\documentclass[%
aip,
jcp,%
amsmath,amssymb,
reprint,%
longbibliography,
floatfix]{revtex4-2}

\usepackage{graphicx} 
\usepackage{hyperref}
\hypersetup{colorlinks=true,
	linkcolor=magenta,
	allcolors=magenta}
\usepackage[utf8]{inputenc}
\usepackage[outercaption]{sidecap} 
\sidecaptionvpos{figure}{c}
\usepackage{newtxtext}
\usepackage{newtxmath}
\usepackage[version=4]{mhchem}
\usepackage{physics}
\usepackage{siunitx}

\usepackage{float}
\usepackage[caption = false]{subfig}
\AtBeginDocument{\RenewCommandCopy\qty\SI}

\newcommand{\SCT}{\ce{{}^SCT}}
\newcommand{\TCT}{\ce{{}^TCT}}
\newcommand{\TBD}{\ce{{}^TBD^$*$}}
\newcommand{\SBD}{\ce{{}^SBD^$*$}}
\newcommand{\TAn}{\ce{{}^TAn^$*$}}
\newcommand{\BDAn}{\ce{BD-An}}
\newcommand{\GS}{\ce{S_0}}

\begin{document}

\setlength{\abovedisplayskip}{4pt}
\setlength{\belowdisplayskip}{4pt}

\title{Unraveling the mechanisms of triplet state formation in a heavy-atom free photosensitizer}
\author{Thomas P. Fay}
\email{tom.patrick.fay@gmail.com}
\affiliation{Department of Chemistry, University of California Berkeley, CA 94720 USA}
\author{David T. Limmer}
\email{dlimmer@berkeley.edu}
\affiliation{Department of Chemistry, University of California Berkeley, CA 94720 USA}
\affiliation{Kavli Energy Nanoscience Institute, Berkeley, CA 94720 USA}
\affiliation{Chemical Science Division Lawrence Berkeley National Laboratory, Berkeley, CA 94720 USA}
\affiliation{Material Science Division Lawrence Berkeley National Laboratory, Berkeley, CA 94720 USA}

\begin{abstract}

Triplet excited state generation plays a pivotal role in photosensitizers, however the reliance on transition metals and heavy atoms can limit the utility of these systems. 
In this study, we demonstrate that an interplay of competing quantum effects control the high triplet quantum yield in a prototypical boron dipyrromethene-anthracene (BD-An) donor-acceptor dyad photosensitizer, which is only captured by an accurate treatment of both inner and outer sphere reorganization energies. Our \textit{ab initio}-derived model provides excellent agreement with experimentally measured spectra, triplet yields and excited state kinetic data, including the triplet lifetime. We find that rapid triplet state formation occurs primarily via high-energy triplet states through both spin-orbit coupled charge transfer and El-Sayed's rule breaking intersystem crossing, rather than direct spin-orbit coupled charge transfer to the lowest lying triplet state. Our calculations also reveal that competing effects of nuclear tunneling, electronic state recrossing, and electronic polarizability dictate the rate of non-productive ground state recombination. 
This study sheds light on the quantum effects driving efficient triplet formation in the BD-An system, and offers a promising simulation methodology for diverse photochemical systems.

\end{abstract}
\maketitle

\section{Introduction}

Photosensitizers harvest photons and transfer energy to other molecules, enabling new chemistry and photophysics, for applications ranging from photocatalysis,\cite{romero_organic_2016,fukuzumi_selective_2013,hari_photocatalyzed_2013,zhao_triplet_2013} 
bioimaging,\cite{filatov_generation_2017,celli_imaging_2010,kaur_recent_2019} and photon upconversion.\cite{zhou_upconversion_2015,zhao_triplettriplet_2011,singh-rachford_photon_2010,khnayzer_upconversion-powered_2012} 
For photosensitizers to function efficiently, the electronic excitation needs to be generated in high yield and persist for a long time. One strategy to achieve this is to engineer the sensitizer to rapidly convert short-lived singlet excited states that are generated through photoexcitation into triplet excited states through intersystem crossing. Relaxation of triplet excited states to the singlet ground state is spin-forbidden, allowing the excitation  to persist for orders of magnitude longer than in singlet excited states. In many photosensitizers, efficient intersystem crossing is facilitated by the presence of heavy atoms, such as transition metals, which enhance the spin-orbit coupling between singlet and triplet excited states. Recently, a large class of heavy-atom free triplet photosensitizers have been developed, capable of producing long-lived triplet excited states in high yields without the presence of heavy atoms.\cite{peng_developing_2016,nguyen_heavy-atom-free_2021,zhang_recent_2021,zhao_recent_2018,filatov_heavy-atom-free_2020} Understanding how triplet formation happens in these systems is essential for the design of other photocatalysts and photosensitizers. Using explicit molecular simulations of \textit{ab initio} derived models, we reveal the mechanism by which triplet state formation occurs in a molecule made of only light elements.

\begin{figure}[b]
    \centering
    \includegraphics[width=0.485\textwidth]{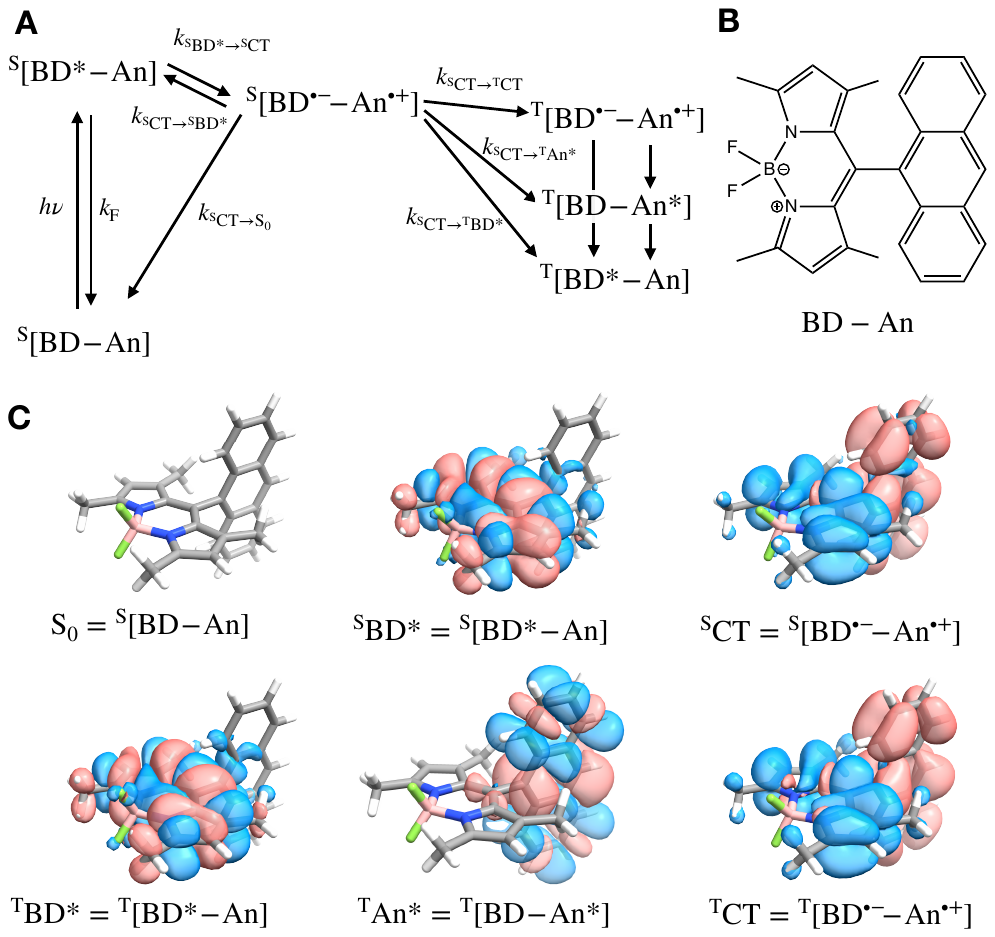}
    \caption{A) Scheme showing the excited state interconversion processes we consider in this work. B) The chemical structure of \ce{BD-An}. C) Difference densities for each of the excited states calculated at with TDDFT using the SOS-$\upomega$B2GP-PLYP functional and def2-TZVP(-f) basis set.}
    \label{fig-scheme}
\end{figure}

In this work we focus on a prototypical heavy-atom-free photocatalyst, the boron-dipyrromethene-anthracene (\BDAn) dyad (chemical structure in Fig.~\ref{fig-scheme}B).\cite{wang_bodipyanthracene_2017,buck_spin-allowed_2019,filatov_control_2018,filatov_generation_2017} \BDAn\ has recently found applications in synthetic chemistry\cite{abuhadba_meso_2021,chen_heavy-atom-free_2023,uddin_twist_2023} and its derivatives have been investigated for phototheraputic applications.\cite{callaghan_structural_2021} The competing photophysical processes and the electronic excited states involved are summarized in Fig.~\ref{fig-scheme}. \BDAn uses excited state charge transfer from an anthracenyl (An) group to the photoexcited \SBD\, forming an \SCT\ state, to enable rapid triplet \TBD\ formation with a high experimental yield, $\Phi_\mathrm{T} = 0.93$--$0.96$\cite{buck_spin-allowed_2019,wang_insights_2019}. Naively one might expect excited state charge transfer to reduce the triplet quantum yield, since the charge transfer state provides a charge recombination pathway for relaxation to the singlet ground state.  However experiments indicate that charge recombination is suppressed by the large charge recombination free energy change, pushing this reverse electron transfer deep into the Marcus inverted regime, where increasing the free energy change increases the activation energy.\cite{buck_spin-allowed_2019} This effect is captured qualitatively by Marcus' theory for the reaction rate constant\cite{marcus_theory_1956,may_charge_2011}
\begin{align}\label{eq-marcus}
    k_{A\to B} = \frac{2\pi |H_{AB}|^2}{\hbar} \frac{1}{\sqrt{4 \pi k_\mathrm{B} T \lambda}} \exp[-\frac{(\Delta A_{A\to B}+\lambda)^2}{4k_\mathrm{B}T \lambda}]
\end{align}
where $H_{AB}$ is the coupling between electronic states $A$ and $B$, $\Delta A_{A\to B}$ is the free energy change of the reaction and $\lambda$ is the reorganization energy, which encodes how solvent fluctuations and intramolecular vibrations control electronic state transitions, $\hbar$ is Planck's constant and $k_\mathrm{B} T$ is Boltzmann's constant times the temperature. Spin conserving charge recombination to the ground state is in the Marcus inverted regime, $-\Delta A_{A \to B} \gg \lambda$, 
which requires a significant activation energy to proceed, whilst for the spin-orbit coupled charge transfer to the triplet excited state $-\Delta A_{A \to B} \approx \lambda$, the reaction is approximately activation-less and thus this spin-forbidden process is competitive, despite $H_{AB}$ being much smaller for the spin-forbidden charge recombination. 
However, Marcus theory is not accurate in the inverted regime due to significant nuclear quantum effects, and alternate triplet formation pathways via high-energy triplet states could contribute, as has been observed in TREPR studies wherein \TCT\ and \TAn\ intermediates were detected at low temperatures.\cite{wang_insights_2019}

We aim to investigate the efficiency of \BDAn\ triplet state generation in solution, going beyond the Marcus picture through first principles computational and theoretical methods, in order to explain how spin-crossover competes with charge recombination and fluorescence in solution. To this end, we interrogate each of the photophysical processes outlined in Fig.~\ref{fig-scheme}A by combining electronic structure calculations, molecular dynamics simulations and non-adiabatic rate theories.\cite{blumberger_recent_2015,may_charge_2011} 
Our aim is to develop models that quantitatively predict experimental observables and give physical insight into mechanisms of triplet formation. We find that effects not captured by Marcus theory, including nuclear tunneling and zero-point energy, have a large effect on the non-adiabatic reaction rate constants, and must be accounted for in our description of these systems.\cite{bader_role_1990,blumberger_recent_2015,lawrence_confirming_2020,blumberger_reorganization_2008} Furthermore, Marcus theory relies on
weak coupling between charge transfer states 
that does not hold for some of the important processes in \BDAn, which we investigate with numerically exact open-system quantum dynamics calculations.\cite{zusman_theory_1988,sparpaglione_dielectric_1988,sparpaglione_dielectric_1988-1,lawrence_calculation_2019}

The importance of solvent effects poses a particular challenge in developing a first principle understanding of triplet state formation, because this necessitates the use of explicit solvent models and molecular dynamics.\cite{blumberger_recent_2015} However common general force fields for organic molecules are only applicable to describe the ground electronic state of these systems. Previous studies have primarily used gas phase electronic structure calculations to rationalize observed behavior,\cite{filatov_control_2018,buck_spin-allowed_2019} but these have not attempted to quantitatively predict rate constants from first principles. To address these challenges, we have developed a protocol for excited state force field parameterization, enabling us to accurately describe solvent fluctuations that control charge transfer processes in ground and excited states. With these tools, we show that the photophysics of \BDAn\ can be quantitatively predicted and mechanisms of triplet formation can be understood in detail. We start by providing 
a brief description of the computational methods used in this study. We then show our results for predicted spectra, free energy changes and rate constants, followed by a discussion of how these can be used to understand efficient triplet formation in \BDAn.


\section{State energies and spectra}
\begin{figure}[t]
    \centering
    \includegraphics[width=0.485\textwidth]{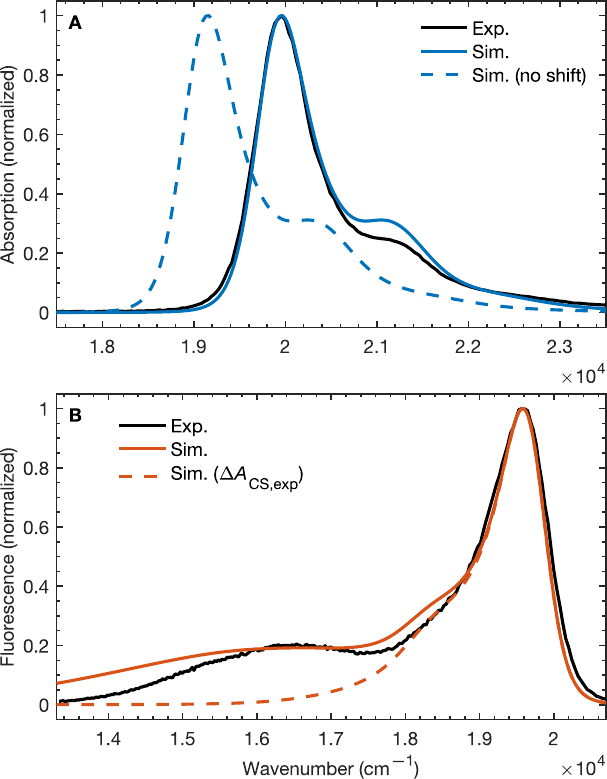}
    \caption{A) Absorption and B) emission spectra of BODIPY-Anth comparing calculated and experimental spectra with and without shifts in the excited state energies. The simulated line-shapes are obtained from the spin-boson mapping described in the main-text with bespoke force-fields for the excited states. The energy differences between excited states were obtained from DLPNO-STEOM-CCSD/def2-TZVP(-f) calculations combined with solvation energies from molecular dynamics. Experimental spectra obtained from Ref.~\onlinecite{buck_spin-allowed_2019}}
    \label{fig-spectra}
\end{figure}

To validate our molecular model, we have computed the \BDAn\ absorption and fluorescence spectra (shown in Fig.~\ref{fig-spectra}). We calculated gas phase energies of the excited states using high-level wave-function based the DLPNO-STEOM-CCSD/def2-TZVP(-f) method\cite{dutta_exploring_2018,dutta_towards_2016} (or DLPNO-CCSD(T)/def2-TZVP(-f) for the \TBD\ and \TAn\ states\cite{guo_communication_2018}), with geometries for each of the excited states obtained from TDA-TDDFT\cite{runge_density-functional_1984,hirata_time-dependent_1999} with the $\upomega$B97X-D3/def2-SVP functional\cite{lin_long-range_2013} and basis set.\cite{weigend_balanced_2005} All calculations were performed with Orca 5.0.3.\cite{neese_orca_2012,neese_software_2018,neese_orca_2020} We found that wave-function based methods, which account for orbital relaxation in the excited state are required in order to obtain an accurate \ce{S_0}-\ce{{}^SCT} gap.

In the absence of solvation effects, the \ce{{}^SBD^$*$} state is lower in energy than the \ce{{}^SCT} state by about \qty{0.5}{eV} (see SI for list of energies), which is inconsistent with the fluorescence spectrum, which shows a clear peak from the CT state at lower energies than the \SBD peak. Thus in order to predict solvation effects and spectral line-shapes, we constructed bespoke force-fields for the ground and excited states of \BDAn, which enabled us to perform  molecular dynamics simulations to efficiently compute spectra with the spin-boson mapping.\cite{wiethorn_beyond_2023} Geometries and Hessians from TDA-TDDFT to were used to parameterize intramolecular force-fields\cite{wang_analytical_2018,yang_multi-state_2013} based on the OPLS-AA force-field.\cite{jorgensen_development_1996,dodda_ligpargen_2017} Electronic polarizability was accounted for using the Drude oscillator model.\cite{halgren_polarizable_2001} We used the same procedure to parameterize both polarizable\cite{halgren_polarizable_2001} and non-polarizable force-fields for the acetonitrile (ACN) solvent, with further non-bonded parameter refinement targeting the dielectric properties of the solvent. The \BDAn\ molecule was solvated in a box of 512 ACN molecules, and energy gap correlation functions were calculated from NVE trajectories, initial after NPT and NVT equilibration (full details are given in the SI). 

From the molecular dynamics (MD) trajectories, the spin-boson mapping was constructed, from which spectra were then calculated.\cite{loco_modeling_2018,zuehlsdorff_optical_2019,wiethorn_beyond_2023} In this approach the full anharmonic potential energy surfaces $V_J$ are mapped onto effective harmonic potential energy surfaces.
Observables of this harmonic model are fully determined by the spectral density $\mathcal{J}_J(\omega)$.
We fit the spectral distribution $\rho_J(\omega) = \mathcal{J}_J(\omega) / \pi\lambda\omega$ from the energy gap correlation function obtained from molecular dynamics,\cite{lawrence_confirming_2020}
\begin{align}
    \rho_J(\omega) = \frac{2 \int_0^\infty \cos(\omega t) \ev{\delta\Delta V (t) \delta\Delta V(0)}_J \dd{t}}{ \pi \ev{\delta\Delta V^2}_J},
\end{align}
where $\Delta V = V_B - V_A$, $\delta\Delta V = \Delta V - \ev{\Delta V}_J$ and $\ev{\cdots}_J$ denotes the classical phase space average over the equilibrium distribution for state $J$ with dynamics calculated on the same surface. For the absorption spectrum we use dynamics on $J = \SBD$ and for the fluorescence spectra we use $J = \GS$ to compute the mapping, and the reorganization energy $\lambda$ is fit from free energy calculations using the same force fields (see below for details). From this mapping the spectra can be calculated from the Fourier transform of correlation function $c_{AB}(t)$,
which is given by
\begin{align}\label{eq-sb-ct}
\begin{split}
	c_{AB}(t)  &= \exp({i({\Delta A_\mathrm{A\to B} }/{\hbar})t - Q'(t) -i Q''(t)}) \\
  Q'(t) &=  \frac{1}{\pi\hbar}\int_0^\infty  \frac{\mathcal{J}(\omega)}{\omega^2}\coth(\frac{\hbar\omega}{2 k_BT}) (1-\cos(\omega t))  \dd{\omega} \\
  Q''(t) &=  \frac{1}{\pi\hbar}\int_0^\infty  \frac{\mathcal{J}(\omega)}{\omega^2} \sin(\omega t) \dd{\omega}.
\end{split}
\end{align}
The absorption, $A_J(\omega)$, and fluorescence, $F_J(\omega)$, spectra (with unit area) are then given by
\begin{align}
    A_{J}(\omega) = \frac{1}{2\pi}\int_{-\infty}^{\infty} e^{i\omega t} c_{\GS,J}(t)\dd{t}\\
    F_{J}(\omega) = \frac{1}{2\pi}\int_{-\infty}^{\infty} e^{i\omega t} c_{J,\GS}(t)^*\dd{t}.
\end{align}
Further details of force-field development and the spin-boson mapping are provided in the SI.

The unshifted spectra calculated from the spin-boson mapping using DLPNO-STEOM-CCSD/def2-TZVP(-f) gas phase energy gaps are shown in Fig.~\ref{fig-spectra} as dashed lines. Our calculated spectra show good overall agreement in the spectral line shapes, without any additional fitting, capturing the narrow \ce{{}^SBD^$*$} peak in the absorption and fluorescence spectra, including a small vibronic side band at about 1500~cm${}^{-1}$ from the main peak, as well as the broad \SCT\ fluorescence band. The agreement in the vibronic structure in the \ce{{}^SBD^$*$} peaks suggests the fitted force fields capture the reorganization energies between excited states relatively well. However we see that the unshifted absorption spectrum calculations underestimates the \SBD energy, which we attribute to the fact that the triple zeta def2-TZVP(-f) basis set is likely still not sufficient for this system. As a result, we shifted all excited states by \qty{805}{cm^{-1}} in order to fit the experimental absorption spectrum. This simple shift is justified by the fact that all excited states shift by $\sim\!0.15$~eV on increasing the basis set size from def2-SVP to def2-TZVP(-f), but differences between excited state energies change by much less (see SI for details). Furthermore it has been found the EOM-CCSD has typical errors of around \qty{0.3}{eV}$\approx$\qty{2400}{cm^{-1}} for charge transfer states, so introducing a shift of \qty{805}{cm^{-1}} seems justifiable. This shift is also used later in the free energy and rate calculations. 

Using the shift from the absorption spectrum, the fluorescence spectrum (Fig.~\ref{fig-spectra}B) was calculated as a weighted sum of the \ce{{}^SCT} and \ce{{}^SBD^$*$} emission spectra, with weights given by the transition dipole moments from DLPNO-STEOM-CCSD, $\mu_{\SBD,\GS}^2 = \qty{7.59}{\text{a.u.}}$ and $\mu_{\SCT,\GS}^2 = \qty{0.54}{\text{a.u.}}$, and equilibrium populations of the two states given by the free energy change of charge separation $\Delta A_{\mathrm{CS}}$, i.e.
\begin{align}
\begin{split}
    F(\omega) &\propto \frac{\mu_{\SBD,\GS}^2}{1+\exp(-\Delta A_{\mathrm{CS}}/k_{\mathrm{B}}T)}F_{\SBD}(\omega) \\
    &+ \frac{\mu_{\SCT,\GS}^2 \exp(-\Delta A_{\mathrm{CS}}/k_{\mathrm{B}}T)}{1+\exp(-\Delta A_{\mathrm{CS}}/k_{\mathrm{B}}T)}F_{\SCT}(\omega)
\end{split}
\end{align}
The assumption of equilibrium between the \SBD\ and \SCT\ states is justified by the fact the time-scale of equilibration of these states is $\sim\! 10^3$ times shorter than the lifetime of these states (as we will discuss shortly). We have also computed the fluorescence spectrum assuming the populations of the \SBD\ and \SCT\ states are given by the experimental estimate, $\Delta A_{\mathrm{CS,exp}}$, based on the approximate Weller equation, which is about \qty{0.2}{eV} larger than our estimate.\cite{buck_spin-allowed_2019} Because $\Delta A_{\mathrm{CS,exp}}>0$, the \SCT\ state is significantly less populated relative to the \SBD\ state and the \SCT\ fluorescence peak is almost completely suppressed, which does not agree with the experimental spectrum. This suggests that the Weller equation cannot be used reliably when free energy changes are close to zero. 
As an interesting aside, the strongest $\SCT-\ce{S_n}$ coupling (see table \ref{tab-free-energies}) is to the \GS\ state, by over a factor of 10, which indicates that the intensity borrowing effect responsible for the \SCT\ emission arises primarily from mixing between \SCT\ and \GS\ states, rather than \SCT\ and \SBD\ states, as has previous been assumed.\cite{wang_insights_2019}

\section{Charge separation and recombination}
\begin{figure*}[t]
    \centering
    \includegraphics[width=0.95\textwidth]{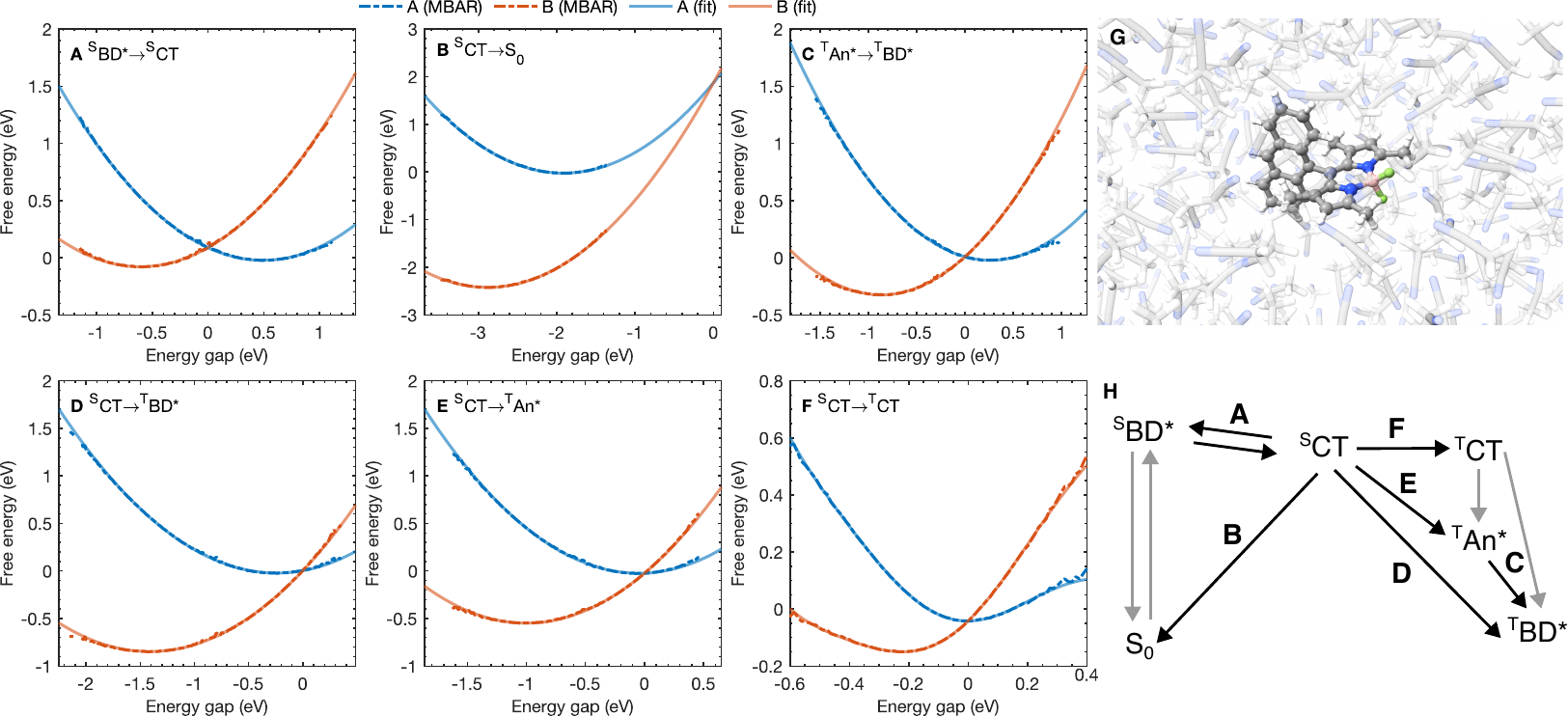}
    \caption{ A--F) Free energy curves for the six \ce{A -> B} processes considered with the reaction \ce{A -> B} labeled on each figure. Points correspond to free energy curves calculated with MBAR and lines correspond to polynomials fitted to the MBAR cumulative distribution functions (see SI for details). G) A snapshot for molecular dynamics simulations on the \GS\ potential energy surface. H) A scheme highlighting the processes in A--F.}
    \label{fig-fec}
\end{figure*}
\subsection{Thermodynamics}
Charge separation, the $\SBD\to\SCT$ process, and charge recombination, the $\SCT\to\GS$ process, both play an important role in efficient triplet formation. Efficient charge separation is required to suppress fluorescence from the \SBD\ state, but slow charge recombination is needed to enable intersystem crossing to occur to generate triplet states. From our excited state force-fields, we have calculated free energy changes associated with these processes from molecular dynamics and the multi-state Bennett acceptance ratio (MBAR).\cite{shirts_statistically_2008,li_understanding_2023} As discussed above, the calculated free energy change for charge separation is \qty{-0.057}{eV}, thus population of the \SBD\ state is reduced and fluorescence is suppressed. 

We have also calculated the rates of these processes from the same MD simulations, by calculating the probability of two states being at resonance. This probability controls the classical Fermi's Golden rule (FGR) rate for the transition between $A$ and $B$.\cite{lawrence_improved_2020} The free energy along the energy gap coordinate, $\Delta V = V_B - V_A$, is related to the energy gap distribution $p_J(\epsilon) = \ev{\delta(\Delta V - \epsilon)}_J$ by\cite{blumberger_recent_2015}
\begin{align}
    A_J(\epsilon) = -k_\mathrm{B} T \ln(p_J(\epsilon)) +(A_B-A_J)
\end{align}
for $J = A$ or $B$. In Fig.~\ref{fig-fec} we show the free energy profiles calculated from MD simulations on each of excited state surfaces with the polarizable ACN model using MBAR. The crossing point of the two curves gives the free energy barrier for the transition, which dictates the classical FGR rate, $k^{\mathrm{class}}_{A \to B} = (2 \pi/\hbar) |H_{AB}|^2 e^{- A_A(\epsilon = 0)/k_\mathrm{B} T} $.\cite{blumberger_recent_2015} If the free energy curve is perfectly quadratic, then this reduces exactly to Marcus theory [Eq.~\eqref{eq-marcus}].\cite{marcus_theory_1956,may_charge_2011,lawrence_confirming_2020} For the charge recombination the crossing point occurs outside of the sampled region, so we extrapolated to the crossing point using a quadratic polynomial \textit{ansatz} for the free energy, fitted to the cumulative distribution function. This procedure was found to result in very little loss in accuracy when compared to umbrella sampling/weighted histogram analysis\cite{kumar_weighted_1992} calculations performed using the non-polarizable ACN model (see SI for details). 

The free energy curves for charge separation and charge recombination are shown in Fig.~\ref{fig-fec}A and B, where we see charge separation lies in the Marcus normal regime, whereas charge recombination is deep in the Marcus inverted regime, with a much larger free energy barrier. Using diabatic state couplings calculated from the generalized Mulliken-Hush method\cite{cave_generalization_1996} with DLPNO-STEOM-CCSD calculations, we can directly calculate the classical FGR rates for these processes (couplings $|H_{AB}|$ are shown in table~\ref{tab-free-energies}). The classical FGR charge separation rate is \qty{4.8e10}{s^{-1}}, about a factor of 10 smaller than the experimentally observed rate of \qty{5.4e11}{s^{-1}}, however the charge recombination rate is predicted to be \qty{1.1d-17}{s^{-1}}, which is more than $10^{24}$ times too small compared to the experimental estimate of $\qty{2.3e7}{s^{-1}}$.\cite{buck_spin-allowed_2019} This enormous discrepancy can be attributed to nuclear quantum effects, in particular the important role of nuclear tunneling in the inverted regime.

\subsection{Quantum effects on rates}

In order to include nuclear quantum effects in the rate calculations, we employed the same spin-boson mapping approach as was used to compute the spectra. The full FGR rate constant is given by
\begin{align}
    k_{A \to B} = \frac{|H_{AB}|^2}{\hbar^2 } \int_{-\infty}^{\infty} c_{AB}(t) \dd{t},
\end{align}
which can be evaluated directly using Eq.~\eqref{eq-sb-ct}. The reorganization energy $\lambda$ is fitted by requiring that the classical limit of the spin-boson mapping reproduces the exact classical limit rate constant, obtained from the classical free energy barrier calculated from MBAR.\cite{li_understanding_2023} This approach to calculating the rate can be regarded as a generalization of the commonly used Marcus-Levich-Jortner theory, accounting for the full frequency dependence of the reorganization energy, which is encapsulated in $\rho_J(\omega)$. The final rate constant is obtained as a simple average over the rate constants calculated with spectral distributions $\rho_A(\omega)$ and $\rho_B(\omega)$.
\begin{table*}[!t]
    {\scriptsize
    \centering
    \begin{tabular}{ccccccc}
         \hline 
          $A$ & $B$ & Calc. $\Delta A_{A\to B}$ (eV)${}^b$  & Exp. $\Delta A_{A\to B}$ (eV)${}^c$ & $\lambda$${}^{c}$ (eV) & $|H_{AB}|$${}^{d}$ (\qty{}{cm^{-1}}) & $k_{A\to B}$${}^{e}$ (\qty{}{s^{-1}})\\
        \hline
        \ce{{}^SBD^$*$} & \ce{{}^SCT}   & \num{-0.057(0.005)} & $+0.13$ &  \num{0.550(0.002)} & 99 & \num{{1.46(0.04)}d11}\\ 
        \ce{{}^SBD^$*$} & \ce{S_0} & \num{-2.4542(0.0004)} & $-2.460$ & \num{8.77(0.08)d-2}${}^{f}$ & -- &  \num{{1.0747(0.0006)}d8}${}^{g}$\\
        \ce{{}^SCT} & \ce{S_0}  & \num{-2.397(0.003)}   & $-2.59$ & \num{0.483(0.003)} & 1904 & \num{{3.4(0.5)}d7}${}^{h}$/\num{{3.6(0.6)}d7}${}^{i}$\\ 
        \ce{{}^SCT} & \TBD & \num{-0.826(0.005)}   & $-0.97$ & \num{0.584(0.001)} & 0.79 & \num{{7.9(0.2)}d7}\\ 
        \ce{{}^SCT} & \TAn & \num{-0.524(0.004)}    & -- & \num{-0.477(0.002)} & 0.63 & \num{{9.7(0.1)}d7}\\ 
        \ce{{}^SCT} & \ce{{}^TCT} & \num{-0.112(0.001)}   & -- &\num{-0.119(0.002)} & 0.21 & \num{{2.86(0.02)}d7}\\ 
        \TAn  & \TBD & \num{-0.302(0.003)}   & -- & \num{0.565(0.002)} & 2.57 &   \num{{1.09(0.01)}d9}\\ 
        \TBD  & \ce{S_0} & \num{-1.638(0.001)}  & $-1.62^j$ & \num{0.512(0.002)}${}^{k}$ & 0.19 & \num{{1.045(0.006)}d4}${}^{k}$
        \\ 
         \hline
    \end{tabular}
    }
    \caption{Uncertainties in the simulated free energy changes and reorganization energies (2$\sigma$) are all $<\qty{0.005}{eV} \approx 0.2\  k_\mathrm{B} T$. 
    ${}^{a}$Free energy changes calculated with non-polarizable ACN, from thermodynamic integration/MBAR. 
    ${}^{b}$Estimated free energy changes from Ref.~\onlinecite{buck_spin-allowed_2019} calculated with the Rehm-Weller equation $\Delta A \approx \Delta G = e (E^{\circ}_\mathrm{D}-E^{\circ}_\mathrm{A}) - \Delta E^* - e^2 / (4\pi\epsilon_0\epsilon_r r_{\mathrm{DA}})$. 
    ${}^{c}$Reorganization energies from equating the $p_J^\mathrm{gaussian}(\epsilon = 0)$ with $p_J(\epsilon = 0)$ (see SI for details). ${}^{d}$ Couplings averaged over gas-phase equilibrium geometries of $A$ and $B$, $|H_{AB}|^2 = (|H_{AB,A}|^2 + |H_{AB,B}|^2)/2$. Details of calculations given in SI. 
    ${}^{d}$Ref.~\onlinecite{buck_spin-allowed_2019}, estimated from spectroscopic measurements. 
    ${}^{e}$Rate constants from spin boson mapping.
    ${}^f$ Linear response value: $\lambda = (\ev{\Delta V}_B - \ev{\Delta V}_A)/2$. 
    ${}^g$Radiative rate constant (Eq.~\eqref{eq-radrate}). 
    ${}^{h}$with recrossing correction and 
    ${}^{i}$without recrossing correction. 
    ${}^j$estimated from spectroscopic measurements.\cite{buck_spin-allowed_2019} 
    ${}^{k}$ Using reorganization energy from non-polarizable umbrella sampling calculations (see SI).
    }
    \label{tab-free-energies}
\end{table*}

\begin{figure}[b!]
    \centering
    \includegraphics[width=0.45\textwidth]{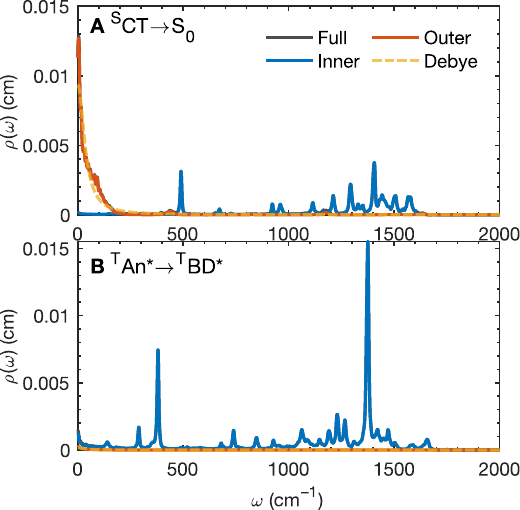}
    \caption{Spectral distribution $\rho(\omega)$ for A) the \SCT$\to$\GS\ process computed from dynamics on the \SCT\ potential energy surface and B) the \TAn$\to$\TBD\ process computed from dynamics on the \TAn\ potential energy surface. The decomposition into inner and outer sphere contributions and the Debye approximation for the outer-sphere component is also shown, $\rho_\mathrm{D}(\omega) = (2/\pi)/(1+\tau_\mathrm{D}^2\omega^2)$, where $\tau_\mathrm{D} = (\epsilon_\infty/\epsilon_{r})\tau_\mathrm{rel}$, and $\tau_\mathrm{rel}$ is the solvent dipole-dipole autocorrelation relaxation time, and $\epsilon_\infty$/$\epsilon_r$ are the optical and static dielectric constants of the ACN model.}
    \label{fig-specden}
\end{figure}

The calculated spectral distributions $\rho_J(\omega)$ can be decomposed into inner sphere, outer sphere and cross-correlated contributions, by decomposing the energy gap into molecular and the remaining environment contributions $\Delta V = \Delta V_\mathrm{mol} + \Delta V_\mathrm{env}$. We find that the cross-correlated contribution is generally negligible for all processes in \BDAn, so the reorganization energy is well-described by a simple sum of inner and outer sphere contributions. The inner and outer sphere spectral distributions are calculated with the non-polarizable ACN/solute model, with the outer sphere contribution scaled down to match the polarizable model outer sphere contributions. As can be seen in Fig.~\ref{fig-specden}A, the low frequency proportion of the spectral distribution for the $\SCT\to\GS$ transition is dominated by the outer sphere contribution arising from solvent molecule fluctuations, making up $\sim 50\%$ of the reorganization energy, which is well approximated by the Debye model.\cite{sparpaglione_dielectric_1988} In contrast, the high frequency region of the spectral density is dominated by the inner sphere contribution from changes in equilibrium bond lengths in the \BDAn\ molecule on charge transfer. The inner sphere spectral distribution has contributions over a range of frequencies from around 500 to \qty{1600}{cm^{-1}}, all of which contribute to tunneling enhancement of the $\SCT\to\GS$ rate, although the dominant mode at $\sim\qty{1400}{cm^{-1}}$ likely corresponds to a \ce{C=C} stretching motion within the aromatic rings. Qualitatively similar spectral distributions were found for the other charge transfer processes. For processes which do not involve charge transfer the spectral distribution is dominated by the inner sphere contribution, as can be seen for the $\TAn\to\TBD$ process in Fig.~\ref{fig-specden}B.

When accounting for nuclear quantum effects, the $\SBD\to\SCT$ rate goes up by a factor of $\sim\!3$ to \qty{1.46e11}{s^{-1}}, and the $\SCT\to\GS$ rate goes up by over $10^{24}$ to \qty{1.0e8}{s^{-1}}, and both calculated rates are now much closer to the experimentally measured values, agreeing much better with the experimental value. Application of Marcus-Levich-Jortner theory with the same inner and outer sphere reorganization energies and a characteristic inner-sphere frequency of \qty{1500}{cm^{-1}} also predicts about a $10^{24}$-fold increase in the rate constant, compared to Marcus theory. This suggests that the large increase is robust to the details of the spectral density. Electronic polarizability is essential to account for in calculating the charge recombination rate. When a non-polarizable model is used instead, the free energy change of the reaction is effectively unchanged but the reorganization energy goes up by nearly \qty{0.1}{eV}. This lowers the activation energy and accelerates the rate by around a factor of three.

\begin{figure}[t]
    \includegraphics[width=0.42\textwidth]{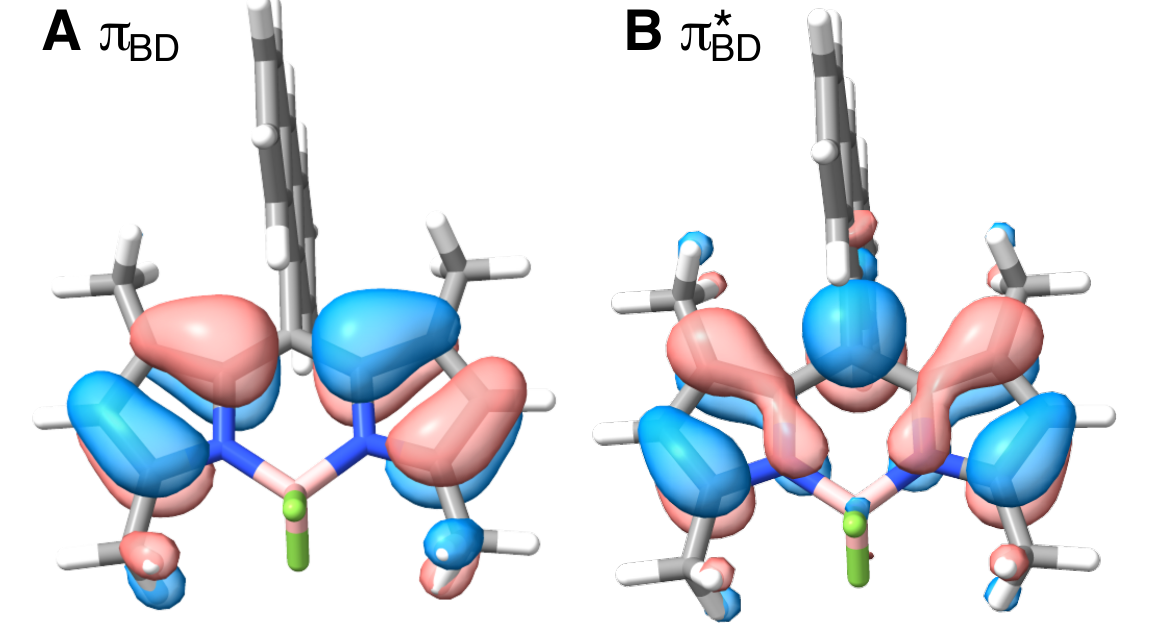}
    \caption{BD orbitals involved in charge separation and recombination A) $\uppi_\mathrm{BD}$ and B) $\uppi_\mathrm{BD}^*$, calculated with $\upomega$B97X-D3/def2-TZVPP/CPCM(ACN) at the \GS\ equilibrium geometry.}\label{fig-bdpi}
\end{figure}

Care should however be taken when using FGR to calculated the charge recombination rate. This is because the diabatic coupling for charge recombination process, $H_{AB} = \qty{1904}{cm^{-1}}$, is about 20 times larger than $k_\mathrm{B}T$, and thus higher-order diabatic coupling effects beyond FGR, may be important (although large nuclear quantum effects in the FGR rate have been observed to reduce the importance of higher order effects).\cite{lawrence_calculation_2019} The large difference in couplings arises from the BD $\uppi$ orbitals involved in the transitions. The $\SBD\to\SCT$ coupling involves an interaction between $\uppi_{\ce{An}}$ and $\uppi_{\ce{BD}}$ (Fig.~\ref{fig-bdpi}A) orbitals, whereas $\SCT\to\GS$ coupling involves the $\uppi_{\ce{An}}$ and $\uppi_{\ce{BD}}^*$ (Fig.~\ref{fig-bdpi}B) orbitals. As can be seen in Fig.~\ref{fig-bdpi} the $\uppi_{\ce{BD}}$ has minimal density on the carbon atom bonded to the An, group, whereas the $\uppi_{\ce{BD}}^*$ orbital does. In order to investigate the potential role of higher-order diabatic coupling effects in the \SCT$\to$\GS\ transition, we have performed Hierarchical Equations of Motion (HEOM) calculations a simple model for this transition. The spectral density for the transition is coarse-grained down to a low-frequency outer-sphere portion described with a Debye spectral density and the inner sphere portion is described with a single under-damped Brownian oscillator spectral density, with a characteristic frequency of \qty{1400}{cm^{-1}}. The coarse-grained spectral density is shown in Fig. A. For this simplified model the exact open quantum system dynamics can be obtained using the HEOM method, and from this the rate constant as a function of $H_{AB}$ can be obtained. These rates are shown in Fig. B. We see that the rate constant is still fortuitously very well described by Fermi's Golden rule for this model, with only a factor of $\sim\!0.9$ reduction in the rate constant at the calculated value of $H_{AB}$. We include this as a correction to the Fermi's Golden rule $k_{\SCT\to\GS}$ that we calculated with the full spectral density. 

\begin{figure}[b]
    \includegraphics[width=0.42\textwidth]{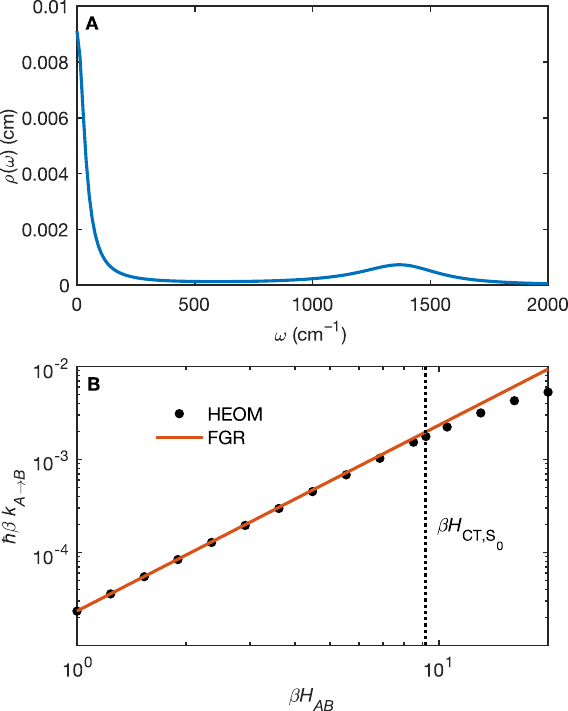}
    \caption{A) The coarse-grained model spectral distribution for the \SCT$\to$\GS\ transition, consisting of a low frequency Debye contribution $\rho_\mathrm{D}(\omega) = (1/2\pi)/(1+(\omega/\omega_D)^2)$, with $\beta\omega_\mathrm{D} = 0.1831$, and an under-damped Brownian oscillator contribution $\rho_{\mathrm{BO}}(\omega) = (1/2\pi) \gamma\Omega^2/((\omega^2 -\Omega^2)^2 + \gamma^2\omega^2)$ with $\beta\gamma = 4$ and $\beta\Omega = 6.76$. The reorganization energy for the Brownian oscillator portion is $\beta \lambda = 8.6780$ and for the Debye portion is $\beta\lambda = 10.1459$. B) The rate constant from HEOM calculations for the coarse-grained spectral density as a function of $H_{AB}$ together with the FGR predictions. The value of $H_{AB}$ for the \SCT$\to$\GS\ transition is also indicated. Calculations were performed using the \texttt{heom-lab} code\cite{fay_simple_2022} using the HEOM truncation scheme from Ref.~\onlinecite{lindoy_quantum_2023}.}\label{fig-heom}
\end{figure}

Radiative recombination from the \SCT\ state can also occur in \BDAn, either through thermally activated delayed fluorescence via the \SBD\ state, or directly. The radiative rates can be calculated from the fluorescence spectra obtained from the spin-boson mapping as\cite{may_charge_2011,fay_coupled_2022}
\begin{align}\label{eq-radrate}
    k_{\mathrm{F},A\to B} = \frac{\mu_{AB}^2}{3\pi \epsilon_0 \hbar c_0} \int_0^\infty \omega^3 F(\omega) \dd{\omega}
\end{align}
where $F(\omega)$ is the fluorescence line-shape computed from the spin-boson mapping. From this we find the fluorescence rate from the \SBD\ state to be \qty{1.1e8}{s^{-1}} and the fluorescence rate from the \SCT \, state to be \qty{7.7e8}{s^{-1}}. Assuming a pre-equilibrium between the \SBD\ and \SCT\ states, as is justified by the large charge separation rate constant, we find that only 63\% of the \GS\ re-formation occurs by direct non-radiative recombination, with 15\% of recombination events happening by radiative \SCT\ recombination and 22\% occurring via \SBD\ thermally activated delayed fluorescence. 

\section{Triplet state formation and lifetime}

As with the charge separation and charge recombination processes, we have calculated the free energy changes and free energy profiles for the three triplet formation pathways: from the \SCT\ state to the \TCT, \TAn\ and \TBD\ states (Fig.~\ref{fig-fec}D--F). Free energy calculations reveal that the three pathways are thermally accessible, with all three states lying lower in energy than the \SCT\ state. Furthermore all three pathways are approximately activation-less, which is at first surprising given that each process has a very different free energy change. The \TBD\ pathways has a larger $|\Delta A_{A\to B}|$, than the \TAn\ pathway, but the \TAn\ pathway has smaller inner and outer sphere reorganization energies, so this pathway is also approximately activation-less. The \TCT\ pathway has a very small reorganization energy which is dominated ($\sim\!90\%$) by the inner sphere contribution, at only \qty{0.11}{eV}. This is because the \SCT\ and \TCT\ states have the same orbitals occupied, so the reorganization energy is dictated only by differences in the exchange energy which alters bonds lengths. However the net exchange energy difference between these states is small because the unpaired electrons have low spatial overlap, so overall the reorganization energy is low and this transition is approximately activation-less. Much like the spin-conserving charge separation and charge recombination, about 50\% of the reorganization energies for the charge transfer processes is outer sphere, with the remaining 50\% arising from inner sphere reorganization, although there is a significant range of reorganization energies for the charge transfer processes, from \qty{0.48}{eV} to \qty{0.58}{eV}. In contrast, the reorganization energies of processes which do not involve charge transfer are dominated by the inner sphere contribution, 89\% for the $\SCT\to\TCT$ spin-crossover and 99\%  for the $\TAn\to\TBD$ triplet-triplet energy transfer, as illustrated in Fig.\ref{fig-specden}B. The triplet-triplet energy transfer still has a reorganization energy comparable to the charge transfer processes, at \qty{0.57}{eV}, due to a large change in bond order in both the BD and An units in this process. Further analysis of the inner/outer sphere reorganization energies are given in the Supporting Information together with all calcualted spectral densities.

We have also calculated the SOC couplings between the different \SCT\ and triplet states using TDDFT ($\upomega$B97X-D3/def2-TZVPP/CPCM(ACN)) and the spin-orbit mean-field (SOMF) treatment of spin-orbit coupling.\cite{neese_efficient_2005,de_souza_predicting_2019} The two spin-orbit coupled charge transfer (SOCT) pathways have the largest SOC couplings, at \qty{0.79}{cm^{-1}} and \qty{0.63}{cm^{-1}} for the \TBD\ and \TAn\, whilst the formally El-Sayed's rule forbidden pathway has a smaller coupling at \qty{0.21}{cm^{-1}}.
Using these couplings and the spin-boson mapping, we find that two El-Sayed's rule allowed transitions, via \TAn\ and \TBD, occur at very similar rates, with $\SCT\to\TBD$ occurring only about 20\% faster than the $\SCT\to\TAn$ formation. The triplet-triplet $\TAn\to\TBD$ energy transfer is also activation-less (Fig.~\ref{fig-fec}C), and has a coupling from fragment energy/charge density (FED/FCD) calculations\cite{yang_multi-state_2013,tolle_electronic_2020,wang_electronic_2021} of \qty{2.57}{cm^{-1}}, and so occurs about 10 times faster than the triplet formation rate, accelerated by a factor of $1.6$ by nuclear quantum effects, so the steady state population of \TAn\ would be difficult to observe spectroscopically at room temperature. The El-Sayed's rule forbidden transition to the \TCT\ state also contributes to triplet formation, although it occurs about 4.5 times slower than \TBD\ formation. The \TCT\ state very rapidly recombines to the \TAn\ or \TBD\ states, with these spin allowed transitions occurring at least $\sim\!10^4$ times faster than the corresponding spin-forbidden transitions, so the \TCT\ state would be very difficult to observe directly at room temperature. Overall the \TCT\, \TAn, and \TBD\ pathways contribute 14\%, 47\%, and 39\% respectively to the overall triplet formation. Surprisingly the most significant pathway is the \TAn\ pathway and not the direct \TBD pathway, which can be rationalized by the lower activation barrier for the \TAn spin-orbit coupled charge recombination. The observation is consistent with TREPR experiments in which all three triplet states were observed, although at much lower temperatures (80K) in a very different medium (a dichloromethane/isopropanol solid matrix). This work shows that multiple triplet formation pathways, including those forbidden by El-Sayed's rule, can contribute at room temperature in polar solvents. The presence of multiple triplet recombination pathways may also explain the large spread of effective spin-orbit coupled charge transfer rates observed in the family of BD-Aryl molecules studied in Ref.~\onlinecite{buck_spin-allowed_2019}. 

\begin{table*}[!t]
    \centering
    {\footnotesize
    \begin{tabular}{ccccccc}
        \hline 
          & $k_{\mathrm{CS,eff}}\ (\text{s}^{-1})$ & $k_{\mathrm{CR,eff}}\ (\text{s}^{-1})$ & $\phi_{\mathrm{CRT}}$ & $\Phi_\mathrm{T}$ & $\Phi_\mathrm{F}$ & $\tau_\mathrm{T} \ (\upmu\text{s})$\\
        \hline
          Calculated & \num{{1.46(0.04)}d11} & \num{{2.31(0.05)}d8} & \num{0.86(0.02)}  & \num{0.80(0.02)}&  \num{0.045(0.01)} & \num{95.7(0.6)}\\
         Experiment [Ref.~\onlinecite{buck_spin-allowed_2019}] & \num{5.4d11} & \num{3.8d8}& \num{0.94} & \num{0.93} & \num{0.01} & -- \\
         Experiment [Ref.~\onlinecite{wang_insights_2019}] & -- &  \num{3.3d8} & 0.98 & \num{0.96}  & \num{0.018} & 78\\
         \hline
    \end{tabular}
    }
    \caption{Calculated and experimental rates, quantum yields and triplet lifetime for the photophysics of \BDAn.}
    \label{tab-rates-yields}
\end{table*}

Using all of the computed rates, we have estimated the observed charge separation and charge recombination rates, as well as the triplet yield. The effective charge separation rate corresponds to the observed equilibration rate between \SBD\ and \SCT\ states i.e. $k_{\mathrm{CS,eff}} = k_{\SBD\to\SCT} + k_{\SCT\to \SBD}$. Likewise the effective charge recombination rate corresponds to the observed decay rate of the \SCT\ state, which under a pre-equilirbium approximation for the \SBD$\rightleftharpoons$\SCT\ interconversion is given by
\begin{align}
    k_{\mathrm{CR,eff}} = p_{\SCT}(k_{\mathrm{CR}} + k_{\mathrm{F},\SCT\to\GS} ) + p_{\SBD} k_{\mathrm{F},\SBD\to\GS} 
\end{align}
where $p_{\SCT} = 1-p_{\SBD} = K_{\mathrm{CS}}/(1+K_{\mathrm{CS}})$, with $K_\mathrm{CS} = e^{-\Delta A_{\SBD\to\SCT}/k_{\mathrm{B}}t}$, $k_{\mathrm{F},\SBD}$ is the calculated fluorescence rate from the \SBD\ state back to the \GS\ state and $k_{\mathrm{CR}}$ is the total recombination rate from the \SCT\ state, i.e. 
\begin{align}
    k_{\mathrm{CR}} = k_{\SCT \to \GS} + k_{\SCT \to \TCT} + k_{\SCT \to \TAn} + k_{\SCT \to \TBD}.
\end{align}
The triplet quantum yield $\Phi_\mathrm{T}$ is calculated as $\Phi_\mathrm{T} = p_{\SCT}(k_{\SCT\to\TCT} + k_{\SCT\to\TAn} + k_{\SCT\to\TBD})\tau_{\mathrm{CR}}$, with $\tau_{\mathrm{CR}} = 1 / k_{\mathrm{CR,eff}}$, and the fluorescence yield $\Phi_\mathrm{F}$ is $\Phi_\mathrm{F} = p_{\SBD} k_{\mathrm{F},\SBD}\tau_{\mathrm{CR}}$. We also computed the fraction of non-radiative transitions which produce a triplet state, $\phi_{\mathrm{CRT}} = \Phi_{\mathrm{T}} / (1-\Phi_\mathrm{F})$, as measured in Ref.~\onlinecite{buck_spin-allowed_2019}. 

The calculated and experimental values of the rates and yields are summarized in table \ref{tab-rates-yields}. Overall we see excellent agreement between the calculated rates/yields and the experimental measurements from Refs.~\onlinecite{buck_spin-allowed_2019,wang_insights_2019}, with less than a factor of $4$ error in the charge separation rate and only a factor of $\sim\!1.6$ error in the charge recombination rate. Similar we only slightly underestimate the triplet yield, with our calculations yielding 0.80, compared to the experimental measurements between 0.93 and 0.96. If we only included the dominant $\SCT\to\TBD$ triplet formation pathway, the triplet quantum yield would only be $\sim\!0.6$, and the error in the rate would be over a factor of 3. We also find that suppression of the charge recombination also plays a large role in efficient triplet formation, which is facilitated by polarizability and recrossing effects. Without including electronic polarizability, the charge recombination rate would be enhanced to $\sim\qty{1.0e8}{s^{-1}}$, which would reduce the triplet quantum yield to $\sim\!0.63$. 
This corroborates the conclusions drawn in Ref.~\onlinecite{buck_spin-allowed_2019}, although we find that multiple triplet pathways also enable the triplet formation to compete with charge recombination, which is suppressed by several effects. 
The net fluorescence quantum yield from \SBD\ that we calculate, 0.045, is also in good agreement with the experimental values, between 0.01 and 0.018. These results suggest that the intersystem crossing rates are being slightly underestimated by our models, possibly due to errors in the reorganization energies or the spin-orbit couplings obtained from TD-DFT, which are all less than \qty{1}{cm^{-1}}.

The triplet lifetime $\tau_\mathrm{T} = 1/k_{\TBD\to\GS}$ plays an important role in determining the utility of a triplet sensitizer or photocatalyst, with longer-lived triplet states allowing more time for diffusive encounters with other molecules enabling more efficient energy transfer. We have also calculated the triplet lifetime for \BDAn using the methods described above, and we also find good agreement between our calculated value for $\tau_\mathrm{T}$ and experimental measurements (table \ref{tab-rates-yields}), with an error of only $\sim\!20\%$. From a simulation perspective, this requires an accurate calculation of the free-energy barrier, which requires enhanced sampling since the transition is very deep in the Marcus inverted regime, since it displays a highly non-quadratic free energy curve. This was achieved using the non-polarizable model with umbrella sampling\cite{torrie_nonphysical_1977} on the energy gap coordinate $\Delta V$ sampled with the Fast-Forward Langevin algorithm.\cite{hijazi_fast-forward_2018} Use of the non-polarizable model is justified because over 99\% of the reorganization energy is inner sphere for both ACN models, and solvent polarizability has less than a \qty{1}{meV} effect on the free energy of the \TBD\ state. As with the spin-conserving charge recombination, because the transition is deep in the inverted regime and the spectral distribution is dominated by high frequency inner sphere contributions, there is a very large nuclear quantum effect of over $10^7$ in the rate constant. One significant source of uncertainty in this is the validity of the spin-boson mapping, where rates calculated from the spectral distribution obtained from \TBD\ and \GS\ dynamics vary by about 50\%. This means that methods that more rigorously account for asymmetry and anharmonicity in the potential energy surfaces, while also accounting for nuclear quantum effects, may be needed to more accurately compute triplet lifetimes for this system and other related systems.\cite{lawrence_improved_2020,trenins_nonadiabatic_2022,heller_instanton_2020} However given the simplicity of the spin-boson mapping and its accuracy in this case, it is clearly still useful in prediction of non-adiabatic rates.

\section{Concluding remarks}

Through this study, we have found that triplet formation in the photosensitizer \BDAn\ hinges on a subtle balance of effects. Firstly charge separation occurs efficiently, which suppresses radiative decay from the \SBD\ state. 
Secondly multiple triplet recombination pathways can operate, due to the range of reorganization energies and free energy changes associated with the rate-limiting intersystem crossing steps in each pathway, and in fact the high-lying triplet pathways make-up the major contribution to triplet formation, rather than direct SOCT to the ground triplet state. Thirdly, spin-conserving charge recombination to the \GS\ state is slowed down a high free energy barrier, with the transition being deep in the inverted regime, as well as diabatic recrossing effects, a significant portion of which arises due to electronic polarizability. The \SCT\ state energy plays an important role in triplet formation, since an increase in energy would increase fluorescence from \SBD, but a decrease in its energy would reduce the barrier for spin-conserving charge recombination because this process is in the Marcus inverted regime. Capturing all of these effects depends on a complete description of the photophysics including accurate calculations of electronic state couplings, explicit solvent fluctuations, polarizability, to capture outer sphere reorganization energies, as well as an accurate description of molecular potential energy surfaces and inner sphere contributions to reorganization energies, as well as the nuclear quantum effects arising due to high frequency vibrations, which accelerate some processes by many orders of magnitude. Enhanced sampling techniques are also necessary to obtain accurate free energy barriers for important processes, namely the triplet decay. 

The simulation techniques and bespoke force-field parametrization approach developed here paves the way for a quantitative modeling of other triplet photosensitizers and related systems,\cite{tong_charge_2020} possibly even enabling straightforward computational screening for properties such as the triplet lifetime. Comparison between simulated and experimental optical spectra indicates that a major source of error is in gas phase energies of excited states. Even the popular wave-function-based DLPNO-STEOM-CCSD method appears to significantly underestimate transition energies, although the ground-state DLPNO-CCSD(T) method which can be used to calculate the \ce{T_1}--\ce{S_0} gap seems robust. We also note that whilst the approximate spin-boson mapping seems fairly reliable for these systems, its application to deep inverted regime processes requires scrutiny. Thus \BDAn could provide an interesting test-bed for recently developed approaches to calculating non-adiabatic transition rates applicable to high-dimensional anharmonic systems.\cite{lawrence_analytic_2018,lawrence_calculation_2019,lawrence_improved_2020,trenins_nonadiabatic_2022,heller_semiclassical_2020,ansari_instanton_2022,heller_instanton_2020,mulvihill_road_2021,mannouch_mapping_2023,runeson_multi-state_2023,lawrence_recovering_2023,amati_quasiclassical_2022,kelly_accurate_2015,runeson_exciton_2023} The $\SCT\to\GS$ transition poses a particular challenge, since it is deep in the inverted regime, nuclear quantum effects are very large and strong diabatic coupling means there may be some effects missed by FGR, which we have estimated using open quantum dynamics simulations. Furthermore in this study we have neglected non-Condon effects\cite{skourtis_protein_2005} and potential spin-vibronic effects,\cite{penfold_spin-vibronic_2018} which could also play a role in determine the rates of conversion between excited states in this system. Future investigations into these potential effects could provide further insight into triplet formation in \BDAn.

Overall, we believe the mechanistic insights gained from this study, which would be difficult to probe directly with experiment alone, could help light the path towards the development of novel and interesting photochemistry in related systems. The observation that high-energy triplet pathways dominate at room temperature opens the door to the intriguing possibility of engineering triplet anti-Kasha's rule systems,\cite{demchenko_breaking_2017} in which higher energy triplet states could be used to drive photochemistry. This could be particularly promising since triplet-triplet energy transfer is strongly distance dependent,\cite{strieth-kalthoff_energy_2018} so spatial separation of chromophore units could be used to extend the lifetime of high-lying triplet states. 
In summary, our comprehensive study highlights the intricate balance of factors influencing triplet formation, including the significance of charge separation efficiency, multiple recombination pathways, and nuclear quantum effects. Moving forward, this mechanistic understanding could steer the development of novel photochemical systems, with a wide range of potential applications.

\subsection*{Author contributions}

 T.P.F.: Conceptualization, Data curation, Formal analysis, Investigation, Software, Writing – original draft, Writing - review \& editing (equal). 
 D.T.L.: Funding acquisition, Supervision, Writing - review \& editing (equal).

\subsection*{Conflict of interest}

The authors declare no conflict of interest.

\subsection*{Acknowledgments}

We would like to thank Tomoyasu Mani for providing data for the experimental absorption and emission spectra for \BDAn\ and for his comments on the manuscript. We would also like to thank Zachariah Page for a useful discussion, and his suggestion of triplet anti-Kasha's rule photochemistry. T.P.F. and D.T.L. were supported by the U.S. Department of Energy, Office of Science, Basic Energy Sciences, CPIMS Program Early Career Research Program under Award DE-FOA0002019.

\subsection*{Data availability}

OpenMM force field files and example scripts to run energy gap calculations, together with initial geometries, can be found at \url{https://doi.org/10.5281/zenodo.10719345}. Other data that is not available in the manuscript or supplementary information is available from the corresponding authors upon a reasonable request.

\subsection*{Supplementary Information}
Details on electronic structure calculations and bespoke force-field parameterization. Details of rate constant calculations. MD simulation details. Supplemental tables of solvent model properties, gas phase state energies, and reorganization energies. OpenMM force field files can be found at \url{https://doi.org/10.5281/zenodo.10719345}.


\subsection*{Glossary of abbreviations}


\begin{tabular}{p{0.28\linewidth}p{0.68\linewidth}}
    ACN & Acetonitrile \\
    An & Anthracene \\
    BD & BODIPY, boron dipyrromethane \\
    CPCM & Conductor-like polarizable continuum \\
    CR & Charge recombination \\
    CS & Charge separation \\
    CT & Charge transfer \\
     DLPNO-STEOM-CCSD & Domain local pair natural orbital similarity transformed equation of motion coupled cluster singles and doubles \\
     DLPNO-CCSD(T) & Domain local pair natural orbital coupled cluster singles and doubles with perturbative triples \\
     EOM-CCSD & Equation of motion coupled cluster singles and doubles \\
     FGR & Fermi's golden rule \\
     HEOM & Hierarchical equations of motion \\
     MBAR & Multi-state Bennett acceptance ratio \\
     MD & Molecular dynamics \\
     NPT & Constant particle number/pressure/temperature molecular dynamics \\
     NVE & Constant particle number/volume/energy molecular dynamics \\
     NVT & Constant particle number/volume/temperature molecular dynamics \\
     WHAM & Weighted histogram analysis \\
     SOCT & Spin-orbit coupled charge transfer \\
     TDA & Tamm-Dancoff approximation \\
     TDDFT & Time dependent density functional theory \\
     TREPR & Time resolved electron paramagnetic resonance
\end{tabular}

\section*{References} 

\bibliography{main-bib.bib}

\end{document}


\setcounter{figure}{0}
\renewcommand{\figurename}{Fig.}
\renewcommand{\thefigure}{S\arabic{figure}}
\setcounter{table}{0}
\renewcommand{\tablename}{Table}
\renewcommand{\thetable}{S\arabic{table}}
\setcounter{equation}{0}
\renewcommand{\theequation}{S\arabic{equation}}

\title{Supplementary Information: Unraveling the mechanisms of triplet state formation in a heavy-atom free photosensitizer}
\author{Thomas P. Fay}
\email{tom.patrick.fay@gmail.com}
\affiliation{Department of Chemistry, University of California Berkeley, CA 94720 USA}
\author{David T. Limmer}
\affiliation{Department of Chemistry, University of California Berkeley, CA 94720 USA}
\affiliation{Kavli Energy Nanoscience Institute, Berkeley, CA 94720 USA}
\affiliation{Chemical Science Division Lawrence Berkeley National Laboratory, Berkeley, CA 94720 USA}
\affiliation{Material Science Division Lawrence Berkeley National Laboratory, Berkeley, CA 94720 USA}
\email{dlimmer@berkeley.edu}
\maketitle
\tableofcontents

\section{Methods and computational details}

\subsection{Electronic structure}

Diabatic excited state force fields for solution-phase simulations were constructed based on gas phase electronic structure calculations. The geometry of each excited state of interest was first optimized using (TD)DFT at the $\omega$B97X-D3/def2-SVP level of theory. Hessians were computed at these geometries for each of these excited states which were used in the parameterization of the force fields. Energy differences gas phase minimum energy geometries were computed using the wavefunction-based DLPNO-STEOM-CCSD method using the def2-TZVP(-f) basis set, except for the \ce{S_0}--\ce{T_1} gap where the DLPNO-CCSD(T)/def2-TZVP(-f) was used. 

The three singlet diabatic states were computed from the calculated using the multi-state generalization of the GMH method,
wherein the dipole moment operator projected along the charge transfer axis, $\vb*{e}_{\mathrm{CT}}\cdot \hat{\vb*{\mu}}$, is first diagonalized, with its eigenvectors defining diabatic states. The $2\times 2$ sub-block of resulting diabatic Hamiltonian corresponding the the \ce{S_0} and \ce{{}^SBD^$*$} states is then diagonalized to define the final (quasi-)diabatic Hamiltonian at each geometry. The resulting $3\times 3$ quasi-diabatic Hamiltonian matrix has the form
\begin{align}
    \vb{H} = \left(\begin{matrix}
        E_{\ce{S_0}} & 0 & V_{\ce{S_0},\SCT} \\
        0 & E_{\ce{{}^SBD^$*$}} & V_{\ce{{}^SBD^$*$},\SCT} \\
        V_{\ce{S_0},\SCT} & V_{\ce{{}^SBD^$*$},\SCT} & E_{\SCT}
    \end{matrix}\right)
\end{align}
Energies and dipole-moment operators were obtained from the DLPNO-STEOM-CCSD calculations. The charge transfer vector $\vb*{e}_{\ce{CT}}$ is the unit vector in the direction $ \vb*{\mu}_{\ce{CT},\mathrm{ad}} - (\vb*{\mu}_{\ce{S_0},\mathrm{ad}}+\vb*{\mu}_{ \ce{{}^SBD^*},\mathrm{ad}})/2$, where $ \vb*{\mu}_{J,\mathrm{ad}} $ is the dipole moment of adiabatic state $J$. 

For triplet-triplet couplings the mixed fragment-charge density/fragment excitation density method was used to compute diabatic states and spin-orbit couplings.\cite{tolle_electronic_2020} L\"owdin charges and transition charge were computed using TDA-TDDFT $\omega$B97X-D3/def2-TZVP at each of the excited state geometries with CPCM treatment of the ACN solvent. From these the fragment charge and excitation density operators were constructed for the adiabatic states corresponding to the states of interest, which were then diagonalized simultaneously using a Jacobi sweep algorithm.\cite{wang_electronic_2021} Spin-orbit couplings were obtained using the TDDFT $\omega$B97X-D3/def2-TZVPP/CPCM(ACN) using the SOMF method in Orca 5.0.3.\cite{neese_efficient_2005,de_souza_predicting_2019}

\subsection{Force field parameterization}

A classical molecular mechanics description of the inter and intramolecular forces was employed for the different electronic states of BODIPY-Anth. For the intermolecular part, a non-polarizable fixed charge model was used for the electrostatic component together with a Lennard-Jones description of repulsion and dispersion effects. Charges for each state were parameterized by taking an average of CHELPG charges computed for each state at the SOS-$\omega$B2G-GPLY/def2-TZVP(-f) level of theory, with and without CPCM treatment of the acetonitrile solvent as implemented in ORCA.\cite{muddana_sampl4_2014,jorge_what_2023} Lennard-Jones parameters were assumed to be the same for each electronic state and are taken from the OPLS-AA force field with boron parameters taken from Ref.~\onlinecite{doherty_revisiting_2017}. 

The intramolecular forcefield is given as a sum of bonds, angles, torsion dihedrals and improper dihedrals, and bond-bond and bond-angle correlation terms, electrostatics and Lennard-Jones terms,
\begin{align}
    V = V_{\mathrm{bond}} + V_{\mathrm{angle}} + V_{\mathrm{bond-bond}} + V_{\mathrm{bond-angle}} + V_{\mathrm{torsion}} + V_{\mathrm{improper}} + V_\mathrm{el} + V_{\mathrm{LJ}}
\end{align}
The long-range forces are ignored for 1-2 and 1-3 bonded atoms and scaled by 0.5 for 1-4 bonded atoms. Periodic dihedral and improper terms are used, as is standard for the parent OPLS-AA forcefield. Harmonic angle forces are used and bi-linear bond-bond and bond-angle terms were used for all 1-3 bonded triples of atoms,
\begin{align}
    V_{\mathrm{bond-bond}} &= \sum_{b,b'\neq b} k_{b,b'} (r_b - r_{b,0})(r_{b'} - r_{b',0}) \\
    V_{\mathrm{bond-angle}} &= \sum_{a,b} k_{a,b} (\theta_a - \theta_{a,0})(r_{b} - r_{b,0})
\end{align}
The $r_{b,0}$ and $\theta_{a,0}$ terms were set to the QM equilibrium geometry parameters. A Morse potential truncated at fourth order was used to describe bond stretches, 
\begin{align}
   V_{\mathrm{bond}} &= \sum_{b} \frac{1}{2}k_b(r-r_{b,eq})^2 (1 - \alpha_b(r-r_{b,eq}) + \frac{7}{12}\alpha_b^2(r-r_{b,eq})^2  ) 
\end{align}
with the Morse $\alpha$ parameter for each bond parameterized from Hessians at displaced geometries in the ground electronic state. 

These potentials were parameterized by first fitting force-constants and equilibrium bond lengths/angles to the gas phase QM hessian calculated at the $\omega$B97X-D3/def2-SVP equilibrium geometries, then refined bond lengths and bond angles to reproduce the gas phase equilibrium structures. The first step in parameterization after LJ parameters and charges were assigned was to fit intramolecular force field terms were by minimizing 
\begin{align}
    \mathcal{L}_{\mathrm{Hess}} = \sum_{A\geq B} \|\vb{H}_{AB,\mathrm{MM}} - \vb{H}_{AB,\mathrm{QM}}\|^2
\end{align}
where $\vb{H}_{AB}$ denotes the partial hessian block. For this step the bonds are treated as harmonic. First the equilibrium bond lengths and angles are allowed to vary arbitrarily but in a second step they are fixed, with any lengths more than \qty{0.0001}{nm} and angles more than \qty{2.5}{\degree} away from the QM equilibrium geometry values fixed at these boundaries. The equilibrium bond lengths and angles are then refined to minimize
\begin{align}
    \mathcal{L}_{\mathrm{geom}} = \sum_{n = 2,3,4}\sum_{A,B \in 1-n\text{ bonded pairs}} (r_{AB} - r_{AB,\mathrm{QM}})^2.
\end{align}
The improper and dihedral parameters were then refined again minimizing $\mathcal{L}_{\mathrm{Hess}}$, and the geometry refinement was repeated after this. The forcefield obtained with the lowest value of $\mathcal{L}_{\mathrm{geom}}$ was taken. The hessian fitting with $\mathcal{L}_{\mathrm{Hess}}$ with multiple randomly displaced geometries was used to parameterize a re-scaling of the bond force constant $k_b$ and the Morse $\alpha_b$ parameter for the ground-state. The same re-scalings and $\alpha_b$ parameters were then used for the excited states. Symmetry was used to constrain values of the force field parameters for equivalent atoms at each stage of the fitting. 

For the polarizable solute model, Drude oscillators are placed on all non-hydrogen atoms. The atom polarizabilities and Thole parameters are fitted by fitting the molecular polarizability for the groundstate for each equilibroum state geometry from SOS-$\omega$B2GP-PLYP/def2-TZVP(-f) calcualtions. We found that the poalrizability does not change considerably between charged states and singlet/triplet states of the individual components, so we consider this a reasonable approximation. Each atomic polarizability and Thole parameter was regularized vs values fitted to small symmetric molecules, namely \ce{C6H6} for aromatic atoms, \ce{C2H6} for the methyl groups, \ce{CF4} for F and \ce{BF4-} for B. In the fitting the CHELPG charges are simultaneously readjusted to optimally reproduce gas phase dipole and quadrupole moments, as well as the molecular polarizability tensor. In the fitting the charges are also regularised against the gas phase CHELPG values to avoid overfitting. This procedure cannot account for how charge may flow around the molecule when it is placed in a dielectric environment. To account for this, we also added the difference in density-derived Hirshfeld charges between vacuum and acetontrile CPCM calculations at the SOS-$\omega$B2GP-PLYP/def2-TZVP(-f) level of theory to the final re-scaled charges obtained from the polarizability fitting.

\subsection{Solvent models}

Two acetonitrile force field models were parameterized to describe this system: a non-polarizable and a polarizable model. Both models were constructed by first performing hessian fitting to the RI-MP2/cc-pVQZ geometry and to obtain harmonic bond-stretch and angle-bend force constants. For the non-polarizable model Lennard-Jones parameters were taken from the OPLS-AA force field and charges were taken as a mixture of CPCM and gas phase CHELPG charges, $q_A = \lambda q_{A}^{\mathrm{CPCM}} + (1-\lambda) q_{A}^{\mathrm{gas}}$, with the mixing parameter $\lambda$ optimized to reproduce the experimental density at 298~K. 

For the polarizable model a single Drude oscillator was added to the nitrile group carbon atom, with an anisotropic polarizability obtained from a RI-MP2/cc-pVQZ calculation. A modified version of the three-step parameterization procedure  was then followed to parameterize charges, Lennard-Jones $\epsilon$ parameters and $\sigma$ parameters.\cite{nunez-rojas_systematic_2019} First the charges were re-scaled according to $q_A = \lambda q_{A}^{\mathrm{CPCM}} + (1-\lambda) q_{A}^{\mathrm{gas}}$ in order to reproduce the experimental dielectric constant at 298~K. Second, the $\epsilon$ parameters in the Lennard-Jones potential were re-scaled to reproduce the vaporization enthalpy at 298~K, and third $\sigma$ parameters were re-scaled to reproduce the experimental density at 298~K. The resulting force-fields both accurately reproduce these properties of ACN at 298~K, as well as the dielectric relaxation time. 

All fitted free energy curves shown in the main text are fitted by fitting a polynomial approximation to the free energy curves for $A$, $\tilde{A}_{A,n}(\epsilon) = \sum_{k=0}^n a_n \epsilon^n$, and $B$, $\tilde{A}_{B,n}(\epsilon) = \tilde{A}_{A,n}(\epsilon) + \epsilon - \Delta A_{A\to B}$, to to cumulative distribution function,
\begin{align}
    \mathrm{CDF}_J(\epsilon) = \int_{-\infty}^\epsilon \dd{\epsilon'} p_J(\epsilon')
\end{align}
for $J = A$ and $B$. The constraint the the integral of $\tilde{p}_J(\epsilon)= 1$ is also added in the fitting. For the fits to the umbrella-sampled free energy curves we only fit to the $A$ curve.

\subsection{Free energy calculations}

The free energy change for the transformation from electronic state $A$ to $B$ was calculated for the non-polarizable model using thermodynamic integration runs, where the potential is given by $V_\lambda(\vb*{q}) = (1-\lambda)V_A(\vb*{q}) + \lambda V_B(\vb*{q})$. Five MD simulations were performed for each transformation with $\lambda$ values equally spaced between 0 and 1. The Fast-Forward Langevin integrator was used to sample configurations with a time-step of 1 fs.\cite{hijazi_fast-forward_2018} MBAR was used to compute the Helmholtz free energy change $\Delta A_{A\to B}$ from these thermodynamic integration runs. WHAM was used with these runs to compute free energy curves and free energy barriers for each electron transfer process. For the $\ce{{}^SCT}\to\ce{S_0}$ and $\TBD\to\ce{S_0}$ processes the crossing region for the two diabatic states was not sampled in the thermodynamic integration runs so umbrella sampling was performed with the the diabatic energy gap $\Delta V =  V_B(\vb*{q})- V_A(\vb*{q})$ used as a biasing coordinate, and again WHAM was used to obtain free energy curves and the free energy barrier for this electron transfer. All fast-forward Langevin dynamics simulations were performed using an in-house code using the OpenMM 7 C++ library for force and energy evaluations.\cite{eastman_openmm_2017}

The box containing \BDAn\ was set-up with 512 ACN molecules, which was found to provide adequately almost completely decayed dipole-dipole correlations, using Packmol.\cite{martinez_p_2009} Three independent NPT trajectories were run with the \ce{S_0} \BDAn\ force field and the non-polarizable ACN model to find an equilibrated box side length of $d = \qty{3.570}{nm}$. In the thermodynamic integration five equally spaced windows were sampled, with each window sampled for \qty{1}{ns} after \qty{0.1}{ns} of equilibration with a FF-Langevin friction constant of $\gamma = \qty{4}{ps^{-1}}$ and a time-step of \qty{1}{fs}. For umbrella sampling the same simulation parameters and trajectory lengths were used, with an umbrella sampling force constant of $k_U = \qty{0.01}{kJ^{-1}.mol} $ with windows centered at integer multiples of \qty{50}{kJ.mol^{-1}}.

For the polarizable model free energy changes were obtained from MBAR with configurations sampled on each electronic state. Free energy barriers were obtained by fitting free energy curves obtained from MBAR to a polynomial. For the charge transfer steps a quadratic polynomial was used but for other processes anharmonicity plays a larger role and a higher order order polynomial was used. Error bars were obtained by bootstrapping the six independent runs. All simulations were performed using the python API of OpenMM 7 and were analyzed with in-house scripts. For the NVT runs, trajectories were equilibrated for \qty{5}{ps} and sampled for \qty{2}{ns} using the extended Lagrangian approach, with a time-step of \qty{1}{fs}, a friction coefficient of $\gamma = \qty{2}{ps^{-1}}$ for the centers of mass and \qty{20}{ps^{-1}} for the Drude oscillators, with the Drude particle temperature set to \qty{1}{K}. 

\subsection{Spin-boson mapping}

For each non-adiabatic process a spin-boson mapping was constructed where an effective potential for each state is constructed as
\begin{align}
    V_J^{\mathrm{SB}}(\vb*{x}) = \sum_{\alpha} \left(\frac{1}{2}m_\alpha \omega_\alpha^2 x_\alpha^2+\delta_{B,J}c_\alpha x_\alpha\right) + \delta_{B,J}(\Delta A_{A\to B}+\lambda) 
\end{align}
The reorganization energy $\lambda$ was fitted to reproduce the free energy barrier to reach the crossing point for $A\to B$, as obtained from the free energy calculations. The rate constants and spectra for the spin-boson mapping depend only on the spectral density
\begin{align}
    \mathcal{J}(\omega) = \frac{\pi}{2}\sum_{\alpha} \frac{c_\alpha^2}{m_\alpha\omega_\alpha}\delta(\omega - \omega_\alpha).
\end{align}
This can be parameterized directly from the Fourier transform of the energy gap correlation function obtained from classical MD. In what follows, we parameterize this as 
\begin{align}
    \mathcal{J}(\omega) = \pi \lambda\omega  \rho(\omega)
\end{align}
where the spectral distribution $\rho(\omega)$ is calculated from MD as
\begin{align}
    \rho_J(\omega) = \frac{\frac{2}{\pi} \int_0^\infty \cos(\omega t) \ev{\delta\Delta V (t) \delta\Delta U(0)}_J \dd{t}}{ \ev{\delta\Delta V^2}_J}.
\end{align}
where $\Delta V = V_B - V_A$, $\delta\Delta V = \Delta V - \ev{\Delta V}_J$ and $\ev{\cdots}_J$ denotes the classical phase space average over the equilibrium distribution for state $J$ with dynamics calculated on the same surface. By fitting the reorganization energy based off of the free energy barrier we ensure the spin-boson mapping we construct becomes exact in the high-temperature limit. There are two possible $\rho(\omega)$ distributions for each transformation, obtained from dynamics on either $A$ or $B$, and in general these distributions will differ. The similarity of the distributions, and in particular the quantities of interest computed from the two different distributions, gives an indication of how accurate the spin-boson mapping can be expected to be. For all computed rate constants, we use an average of rate constants computed with $\rho_A(\omega)$ and $\rho_B(\omega)$, although we have found that first averaging the spectral distributions and computing rates from this averaged distribution yields almost identical final rate constants.

The correlation function appearing in the full FGR rate constant can be evaluated as
\begin{align}
\begin{split}
	&\ \frac{\Tr[e^{-\beta \op{H}_{A}}e^{-i\op{H}_{A} t/\hbar}e^{+i\op{H}_{B} t/\hbar}]}{\Tr[e^{-\beta \op{H}_{A}}]}  \\
	&=\exp \bigg( -i\frac{\Delta E_\mathrm{AB} t}{\hbar} -\frac{1}{\pi\hbar}\int_0^\infty  \frac{\mathcal{J}(\omega)}{\omega^2}\left[\coth(\frac{\hbar\omega}{2 k_BT}) (1-\cos(\omega t)) - i \sin(\omega t) \right] \dd{\omega} \bigg),
\end{split}
\end{align}
where $\Delta E_{AB} = -\Delta A_{A\to B}$. This is evaluated by discretizing the spectral density using the method in Ref.~\onlinecite{lawrence_confirming_2020}. The spectra can also be directly obtained from the Fourier transform of this function. The correlation function is multiplied by an exponential decay $e^{-t/\tau}$ when computing the spectra to account for instrument broadening. A time-constant $\tau = \qty{350}{fs}$ is used, as has been used in other studies.

\begin{figure}
    \centering
    \includegraphics[width=0.65\textwidth]{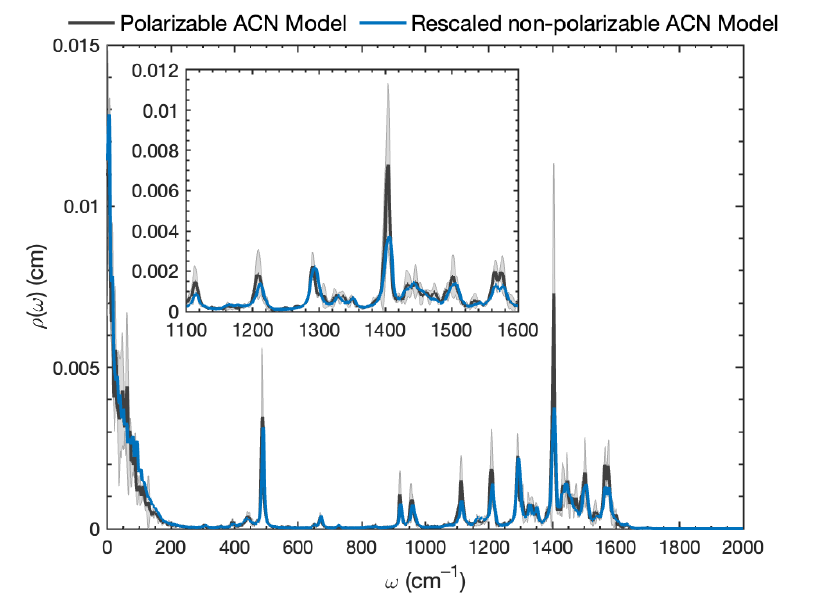}
    \caption{Full polarizable ACN spectral distribution and the approximate form given by Eq.~\eqref{eq-approx-rho} for the $\SCT\to\GS$ process, calculated with dynamics on the \SCT\ surface.}
    \label{fig-scfvsnonpol}
\end{figure}

Spectral densities for the non-polarizable model were obtained from 20 NVE trajectories each \qty{30}{ps} long with a time step of \qty{0.5}{fs}. The cross correlated portion of the spectral distribution was found to have a negligible effect of the calculated rate constants (as long as the total reorganization energy was held fixed at the fitted value). Furthermore, given that the Debye relaxation time for the non-polarizable and polarizable models are almost identical (see table \ref{tab-acn}) and that this model very accurately describes the outer sphere contribution, we constructed the polarizable model spectral distribution as
\begin{align}\label{eq-approx-rho}
    \rho^\mathrm{pol}(\omega) = (1-f_\mathrm{inner}^\mathrm{pol})\rho^\mathrm{non-pol}_\mathrm{outer}(\omega) + f_\mathrm{inner}^\mathrm{pol}\rho^\mathrm{non-pol}_\mathrm{inner}(\omega).
\end{align}
This avoids the need to run expensive self-consistent field Drude oscillator calculations to obtain the spectral density for the non-polarizable model. The proportion of inner sphere reorganization energy was set as
\begin{align}
    f_\mathrm{inner} = \frac{1}{2} \frac{\ev{\Delta V_\mathrm{mol}}_B - \ev{\Delta V_\mathrm{mol}}_A}{\lambda_{\mathrm{LR}}}.
\end{align}
which was found to agree well with values obtained from the co-variance based estimates. In Fig.~\ref{fig-scfvsnonpol} we show the approximated spectral distribution based off of the re-scaled non-polarizable ACN calculations to the full spectral distribution calculated from three NVE trajectories with the full polarizable ACN potential with SCF Drude oscillator integration. We see excellent agreement (within the uncertainty of the polarizable model simulation) between the approximate and full spectral distribution.

The coupling $H_{AB}$ for the rate constant calculation was taken as the root-mean-square of the couplings between the two gas-phase equilibrium geometries of $A$ and $B$, i.e. $H_{AB} = \sqrt{(|H_{AB,A}|^2 + |H_{AB,B}|^2)/2}$. This ensures a symmetric definition of the rate constants, but does not account for non-Condon effects.

\subsection{Hierarchical equations of motion calculations}

The hierarchical equations of motion (HEOM) method, as implemented in the \texttt{heom-lab} Matlab code\cite{heom-lab,fay_simple_2022} was used to obtain exact rate constants for the model spectral density for the \SCT$\to$\GS\ transition. The system was initialised in the state $\hat{\rho}(0)= \dyad{A} \hat{\rho}_\mathrm{b}$ and the long time dynamics of the $B$ state population were fitted to a function
\begin{align}
    p_B(t) = (p_\infty - p_0) (e^{-k t} - 1) + p_0  
\end{align}
in order to obtain the rate constant. The calculations used the HEOM truncation scheme from Ref.~\onlinecite{lindoy_quantum_2023} with $L = 80$, $L\Gamma = 6$ and a Matsurbara expansion was used for the bath correlation functions with the $M=1$ Matsubara mode included explicitly in the calculations. The remaining Matsubara terms were treated with the low temperature correction scheme in Ref.~\onlinecite{fay_simple_2022}. Dynamics were run out to $t = 10\hbar\beta $.

\section{Intensity borrowing}

The effective transition dipole moments used in the spectral calculation are taken directly from the adiabatic DLPNO-STEOM-CCSD calculations. From the diabatization we can analyze the origin of the relatively large transition dipole moment for the \SCT\ state. The adiabatic states are given by first order perturbation theory in the diabatic couplings $H_{AB}$ as
\begin{align}
    \ket{\Psi_A} = \ket{A} + \sum_{J\neq A} \ket{J}\frac{H_{A,J}}{E_A - E_J}
\end{align}
From this we find the mixing coefficient for the \SCT\ and \GS\ states is $0.12$ compared to $0.053$ for the \SCT\ and \SBD\ states. The transition dipole moment is given to lowest order in perturbation theory by
\begin{align}
    \mel{\Psi_A}{\hat{\vb*{\mu}}}{\Psi_B} = \mel{A}{\hat{\vb*{\mu}}}{B} + \sum_{J\neq A} \mel{J}{\hat{\vb*{\mu}}}{B}\frac{H_{A,J}^*}{E_A - E_J} + \sum_{J\neq A} \mel{A}{\hat{\vb*{\mu}}}{J}\frac{H_{B,J}}{E_B - E_J}
\end{align}
and for the $\SCT \to \GS$ transition this is given approximately by
\begin{align}
\begin{split}
     \mel{\Psi_{\SCT}}{ \hat{\vb*{\mu}} }{\Psi_{\GS}} &= \mel{\SCT}{\hat{\vb*{\mu}}}{\GS} 
     + \left(\ev{\hat{\vb*{\mu}}}{\GS} - \ev{ \hat{ \vb*{\mu} } }{\SCT} \right) \frac{ H_{\SCT,\GS} }{E_{\SCT} - E_{\GS} } \\
     &+ \mel{\SBDs}{ \hat{\vb*{\mu}} }{\GS} \frac{ H_{\SCT,\SBDs} }{E_{\SCT}-E_{\SBDs}}.
\end{split}
\end{align}
The second term in this expression, arising from mixing of the \SCT\ and \GS\ states dominates. Thus it is primarily \SCT-\GS\ mixing that enables intensity borrowing in the \SCT\ emission.

\section{Supplementary data}

\subsection{Solvent model properties}

In table \ref{tab-acn} we show a summary of solvent properties calculated from molecular dynamics for the solvent models. All simulations used a box of 512 ACN molecules. All simulations were performed at \qty{298}{K}. These properties were considered as good target properties for accurate simulations of electron transfer in ACN at \qty{298}{K}.

\begin{table}[h]
    \centering
    \begin{tabular}{cccc}
        \hline 
          Property & Non-polarizable & Polarizable (eV) & Experiment \\
        \hline
        Density (\qty{}{g.cm^{-1}}) & \qty{0.7758(0.001)}{} & \qty{0.7779(0.002)}{} & \qty{0.7778}{}\cite{grant-taylor_thermal_1976}\\
        $\Delta H_{\mathrm{vap}}$ (\qty{}{kJ.mol^{-1}}) & \num{39.2(0.3)} & \num{34.2(0.4)} & \num{33.225}\cite{putnam_entropy_1965}\\
        Static dielectric constant, $\epsilon_0$ & \num{29.8(2.7)}& \num{37.5(2.8)} & \num{35.55}\cite{asaki_dielectric_2002}\\ 
        Optical dielectric constant, $\epsilon_\infty = n^2$ & 1 & \num{1.75} & \num{1.81}\cite{moutzouris_refractive_2014} \\
        Dielectric relaxation time, $\tau_\mathrm{rel}$ (\qty{}{ps}) & \num{4.1} & \num{3.0} & \num{3.3}\cite{asaki_dielectric_2002} \\
        Debye time, $\tau_\mathrm{D} = (\epsilon_\infty / \epsilon_{0}) \tau_\mathrm{rel}$ (\qty{}{ps}) & \num{0.141} & \num{0.140} & \num{0.168} \\
         \hline
    \end{tabular}
    \caption{Physical properties of ACN calculated for the polarizable and non-polarizable models and experimental values for comparison.}
    \label{tab-acn}
\end{table}

\subsection{Gas phase energies}

A comparison of gas phase minimum energy geometry energies, calculated with different basis sets, relative to the \GS\ state energy is shown in table \ref{tab-energies}.

\begin{table}[h]
    \centering
    \begin{tabular}{cccc}
        \hline 
          State & def2-SVP (eV) & def2-TZVP(-f) (eV) & Shifted def2-TZVP(-f) \\
        \hline
        \ce{S_0} & 0 & 0 & 0 \\
        \ce{{}^SBD^$*$} & 2.232 & 2.389 &  2.478 \\
        \ce{{}^SCT} & 2.780 & 2.927 & 3.017 \\
        \ce{{}^TCT}${}^a$ & 2.872 & 3.025 & 3.114 \\
        \ce{{}^TAn^$*$}${}^b$ & -- & 2.017 & 2.017 \\
        \ce{{}^TBD^$*$}${}^b$ & 1.593 & 1.558 & 1.558\\
         \hline
    \end{tabular}
    \caption{Gas phase energies of the different excited states of \ce{BD-An} at their respective minimum energy geometries (from gas phase TDA-TDDFT $\upomega$B97X-D3/def2-SVP geometry optimizations) computed at the DLPNO-STEOM-CCSD level of theory with different basis sets. ${}^a$Geometry obtained from calculation with CPCM treatment of ACN solvent. ${}^b$Energies computed from $S=1$ ground state DLPNO-CCSD(T) calculations.}
    \label{tab-energies}
\end{table}

\subsection{Couplings}
\begin{table}[b]
    \centering
    \begin{tabular}{cccc}
        \hline 
        $A$ & $B$ & $|H_{AB}|$ at $A$ (cm$^{-1}$) & $|H_{AB}|$ at $B$ (cm$^{-1}$)\\
        \hline
         \ce{{}^SBD^$*$} & \ce{{}^SCT} & 64${}^{a}$ & 125${}^{a}$ \\ 
        \ce{S_0} & \ce{{}^SCT} & 865${}^{a}$ & 2550${}^{a}$\\ 
        \ce{{}^SCT} & \ce{{}^TCT} & 0.28${}^{a}$ & 0.11${}^{a}$\\ 
        \ce{{}^SCT} & \ce{{}^TAn^$*$} & 0.62${}^{b}$ & 0.64${}^{b}$ \\ 
        \ce{{}^SCT} & \ce{{}^TBD^$*$} & 0.79${}^{b}$ & 0.80${}^{b}$ \\ 
        \ce{{}^TCT} & \ce{{}^TAn^$*$} & 505${}^{c}$ & 351${}^{c}$ \\ 
        \ce{{}^TCT} & \ce{{}^TBD^$*$} & 12${}^{c}$ & 53${}^{c}$ \\ 
        \ce{{}^TAn^$*$} & \ce{{}^TBD^$*$} & 0.23${}^{c}$ & 3.6${}^{c}$ \\ 
        \ce{S_0} & \ce{{}^TBD^$*$} & 0.17${}^{b}$ & 0.21${}^{b}$ \\ 
         \hline
    \end{tabular}
    \caption{Couplings between different states in \BDAn. ${}^{a}$ GMH couplings from DLPNO-STEOM-CCSD/def2-SVP. ${}^{b}$ SOMF couplings from TDDFT $\upomega$B97X-D3/def2-TZVPP/CPCM(ACN) spin orbiot couplings ($H_{AB}^2 = {\sum_{\alpha = x,y,z} |H_{AB,\alpha}|^2}$ where $\alpha $ denotes component of the SOC operator).\cite{neese_efficient_2005} ${}^{c}$ FED/FCD couplings\cite{tolle_electronic_2020} from TDA-TDDFT $\upomega$B97X-D3/def2-TZVP/CPCM(ACN).}
    \label{tab-couplings}
\end{table}

We have computed both the spin-conserving diabatic couplings as well as spin-orbit couplings between singlet and triplet states. Table \ref{tab-couplings} shows the computed values of these $H_{AB}$ at the minimum energy geometries of state $A$ and $B$. The spin-orbit couplings are of course significantly smaller than the spin-conserving diabatic couplings. We note that couplings between the \SCT\ and \ce{S_0} state are significantly larger than the \SCT-\SBD\ couplings, and the SOC couplings are all small, with the coupling for the El-Sayed's rule forbidden \SCT$\to$\TCT transition not being significantly smaller than the other SOC mediated charge transfer processes, despite there not being a change in orbital occupancy between these states. This can likely be attributed to the fact that although the SOC-mediated CT processes involve a change in orbital occupancy, the orbitals involved are localised on different fragments of the molecule, so the poor orbital overlap reduces the SOC between these states. Overall we see that all three \SCT$\to$\ce{T_n} pathways, which are all feasible according to the free energy changes, are weakly allowed by the SOC interaction.

\subsection{Umbrella sampling}

\begin{figure}[h]
    \centering
    \includegraphics[width=0.48\textwidth]{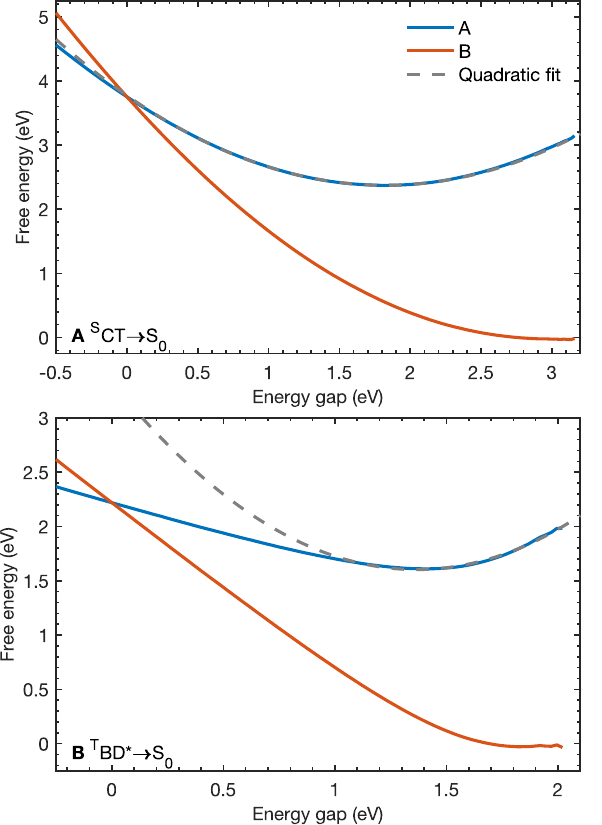}
    \caption{Free energy curves for the \SCT$\to$\ce{S_0} (A) and \TBD$\to$\ce{S_0} (B) processes obtained from umbrella sampling on the \SCT\ and \TBD\ surfaces. The dashed line shows a quadratic fit to the curves with data c at energy gaps $>\qty{1}{eV}$, i.e. close to the $A$ minimum.}
    \label{fig-umbrella}
\end{figure}

For the \SCT$\to$\ce{S_0} and \TBD$\to$\ce{S_0} processes the transitions lie very deep in the Marcus inverted regime, which means that enhanced sampling is required to obtain accurate free energy profiles. We have performed Umbrella sampling directly on the energy gap coordinate $\Delta V$ for these two transitions, in order to sample rare fluctuations of the solvent and molecule in which $\Delta V = 0$ and subsequently compute the free energy barrier. Efficient molecular dynamics sampling on the biased potentials with the polarizable ACN model is not possible, because the extended Lagrangian approach cannot be applied, so we only use the non-polarizable ACN model in these simulations. In Fig.~\ref{fig-umbrella} we show these free energy curves, together with a quadratic fit to the $A$ state curve data (obtained by fitting the corresponding cumulative distribution function, see SI for details), excluding the portion of the curve where the energy gap, $\epsilon$, is less than \qty{1}{eV}. We see that in the \SCT$\to$\ce{S_0} (Fig.~\ref{fig-umbrella}A) the quadratic fit provides an excellent fit, which extrapolates accurately to the crossing point. The effective fitted Marcus reorganization energy, $\lambda_\mathrm{fit}$, obtained by equating the Marcus-theory (Gaussian) probability density at the crossing point and the Umbrella sampling probability density, agree very well, with $\lambda_{\mathrm{fit}} = \qty{0.590}{eV}$ compared to the quadratic estimate $\lambda_{\mathrm{quadratic}} = \qty{0.585}{eV}$. This suggests that obtaining $\lambda_{\mathrm{fit}}$ for the polarizable model by fitting the free energy curves (which do not sample the $\epsilon = 0$ region) to a quadratic and extrapolating should introduce an error of only $\simeq\!\qty{0.005}{eV}$. In stark contrast, for the \TBD and \ce{S_0} states there is a large deviation between a quadratic fit and the computed free energy curves for the \TBD$\to$\ce{S_0} process (Fig.~\ref{fig-umbrella}B). This can be attributed to the fact that $\sim\! 98\%$ of the reorganization energy for the \TBD$\to$\ce{S_0} arises from intra-molecular inner sphere contributions (see below for how this is calculated), for which asymmetry in the vibrational frequencies and anharmonicity in the molecular potential energy surfaces plays a significant role. However for the \SCT$\to$\ce{S_0} process the inner sphere contribution is only about 40\% of the total reorganization energy, so the outer sphere contribution arising from fluctuations in the solvent polarization dominates, which is very well described by a Gaussian field theory.

\subsection{Reorganization energies}

It is interesting to quantify how different the reorganization energies are between the different state transitions, but as observed above, not all of the free energy curves are perfectly quadratic. However, as stated above, we can fit $p_{A}(\epsilon=0)$ to the Gaussian functional form in Marcus theory $p_{A}^{\mathrm{G}}(\epsilon) = (4\pi\lambda k_\mathrm{B}T)^{-1/2} e^{-(\Delta A_{A\to B}-\epsilon+\lambda)^2/4\lambda k_\mathrm{B}T}$, to obtain a effective reorganization energy $\lambda_\mathrm{fit}$. Predictably it is smallest for the \SCT$\to$\TCT\ transition, which involves no change in orbital occupancy or significant charge redistribution, at \qty{0.11}{eV}. 

In order to further investigate the role of non-Gaussian fluctuations in transitions between different excited states, we have computed several other estimates of the the Marcus theory reorganization energy for each of the processes. In addition to the fitted value, we can also estimate the reorganization energy from linear-response theory as\cite{blumberger_reorganization_2008,lawrence_confirming_2020}
\begin{align}
    \lambda_{\mathrm{LR}} = \frac{1}{2}\left(\ev{\Delta V}_A -\ev{\Delta V}_B\right)
\end{align}
or from the energy gap fluctuations on a given surface $J = A$ or $B$,\cite{blumberger_reorganization_2008,lawrence_confirming_2020}
\begin{align}
    \lambda_{\mathrm{var},J} = \frac{\ev{(\Delta V-\ev{\Delta V}_J)^2}_J}{2k_\mathrm{B}T} = \frac{\ev{\delta\Delta V^2}_J }{2k_\mathrm{B}T}.
\end{align}
If the linear response approximation in Marcus theory is valid, then all of these estimates will agree, and therefore deviations between these estimates are indicative of non-Gaussian effects. We see that for the \SBD$\to$\SCT\ and \SCT$\to$\ce{S_0} the four estimates are in very close agreement, but for the other processes there are more significant deviations between the estimates, although these deviations are still relatively small. The largest deviations are for the processes dominated by the inner-sphere contribution to the reorganization energy, namely the \SCT$\to$\TCT, \TAn$\to$\TBD\ and \TBD$\to$\ce{S_0} processes, which is unsurprising given the greater importance of asymmetric vibrational frequencies and anharmonicity in the inner sphere contribution. We see however that the linear-response estimate for the reorganization energy generally agrees very well with the fitted value, which suggests that asymmetry in the energy gap fluctuations between surfaces $A$ and $B$ contribute primarily to the free energy differences between these states, and they have only a small effect on the barrier height. 

The variance based estimators for the reorganization energy also enable us to decompose the reorganization energy in inner sphere, outer sphere, and inner-outer-sphere cross-correlation contributions. We decompose the energy gap $\Delta V$ into an intramolecular contribution and a remaining environment contribution, $\Delta V = \Delta V_\mathrm{mol} + \Delta V_\mathrm{env}$. We take the molecular contribution to be the difference in energies at a given configuration in the absence of the solvent, and the environment term as the remainder $ \Delta V_\mathrm{env} =  \Delta V_\mathrm{mol} - \Delta V$, which is effectively the difference in solvation energies between the two states $A$ and $B$. This allows us to decompose the reorganization energy as $\lambda_{\mathrm{var},J} = \lambda_{\mathrm{var},J}^{\mathrm{inner}}  +\lambda_{\mathrm{var},J}^{\mathrm{outer}} + 2 \lambda_{\mathrm{var},J}^{\mathrm{cross}}$, with the contributions given by\cite{blumberger_reorganization_2008}
\begin{align}
    \begin{split}
    \lambda_{\mathrm{var},J}^{\mathrm{inner}} &= \frac{\ev{\delta \Delta V_\mathrm{mol}^2}_J}{2 k_\mathrm{B} T} , \
    \lambda_{\mathrm{var},J}^{\mathrm{outer}} = \frac{\ev{\delta \Delta V_\mathrm{env}^2}_J}{2 k_\mathrm{B} T} \\
     &\lambda_{\mathrm{var},J}^{\mathrm{cross}} = \frac{\ev{\delta \Delta V_\mathrm{mol}\delta \Delta V_\mathrm{env}}_J}{2 k_\mathrm{B} T}.
     \end{split}
\end{align}
We see that variations in the total reorganization energy arise from variations in both the inner and outer sphere contributions. For the charge transfer processes the inner sphere contribution is typically slightly smaller than the outer sphere contribution, but for transitions which are not accompanied by charge transfer the inner sphere contribution makes up over $90\%$ of the total reorganization energy. We also see that for all transitions the cross correlation contributions are very small compared to the total reorganization energy. 
\begin{table*}
    \centering
    \begin{tabular}{ccccccccccc}
        \hline 
          & A & B & $\lambda_{\mathrm{fit}}/\text{eV}$ & $\lambda_{\mathrm{LR}}/\text{eV}$ & $\lambda_{\mathrm{var,A}}/\text{eV}$ &$\lambda_{\mathrm{var,B}}/\text{eV}$ & $f^{\mathrm{inner}}_{\mathrm{var}}$&$\lambda_{\mathrm{var}}^{\mathrm{inner}}/\text{eV}$&$\lambda_{\mathrm{var}}^{\mathrm{outer}}/\text{eV}$&$\lambda_{\mathrm{var}}^{\mathrm{cross}}/\text{eV}$\\
        \hline
         Polarizable ACN & $\ce{{}^SBD}^*$ & $\ce{{}^SCT}$ & 0.550 & 0.550 & 0.557 & 0.525 & 0.436 & 0.235 & 0.235 & -0.008   \\
           & $\ce{{}^SCT}$ & $\ce{S_0}$ & 0.483 & 0.482 & 0.464 & 0.471 & 0.453 & 0.212 & 0.212 & 0.001   \\
           & $\ce{{}^SCT}$ & $\ce{{}^TCT}$ & 0.119 & 0.125 & 0.132 & 0.169 & 0.891 & 0.134 & 0.134 & 0.001  \\
           & $\ce{{}^SCT}$ & $\ce{{}^TAn}^*$ & 0.477 & 0.478 & 0.433 & 0.487 & 0.494 & 0.227 & 0.227 & -0.002 \\
           & $\ce{{}^SCT}$ & $\ce{{}^TBD}^*$ & 0.565 & 0.567 & 0.598 & 0.613 & 0.997 & 0.604 & 0.604 & -0.004    \\
            & $\ce{{}^TAn}^*$ & $\ce{{}^TBD}^*$ & 0.584 & 0.584 & 0.547 & 0.618 & 0.507 & 0.294 & 0.294 & -0.011   \\
            & $\ce{{}^TBD}^*$ & $\ce{S_0}$ & -- & 0.225 & 0.333 & 0.206 & 0.967 & 0.261 & 0.261 & -0.002  \\
            \hline
        Non-pol ACN & $\ce{{}^TBD}^*$ & $\ce{S_0}$ & 0.512 & 0.236 & 0.346 & 0.239 & 0.98 & 0.283 & 0.008 & -0.002 \\
        & $\ce{{}^SCT}^*$ & $\ce{S_0}$ & 0.590 & 0.581 & 0.603 & 0.590 & 0.40 & 0.233 & 0.345 & 0.002\\
         \hline
    \end{tabular}
    \caption{Uncertainties in the reorganization energies (2 standard errors in mean obtained from bootstrapping data NVT runs initialized from independent initial configurations) are all $<0.005$ eV.}
    \label{tab-reorg}
\end{table*}

\begin{figure*}[h]
    \centering
    \includegraphics[width=0.85\textwidth]{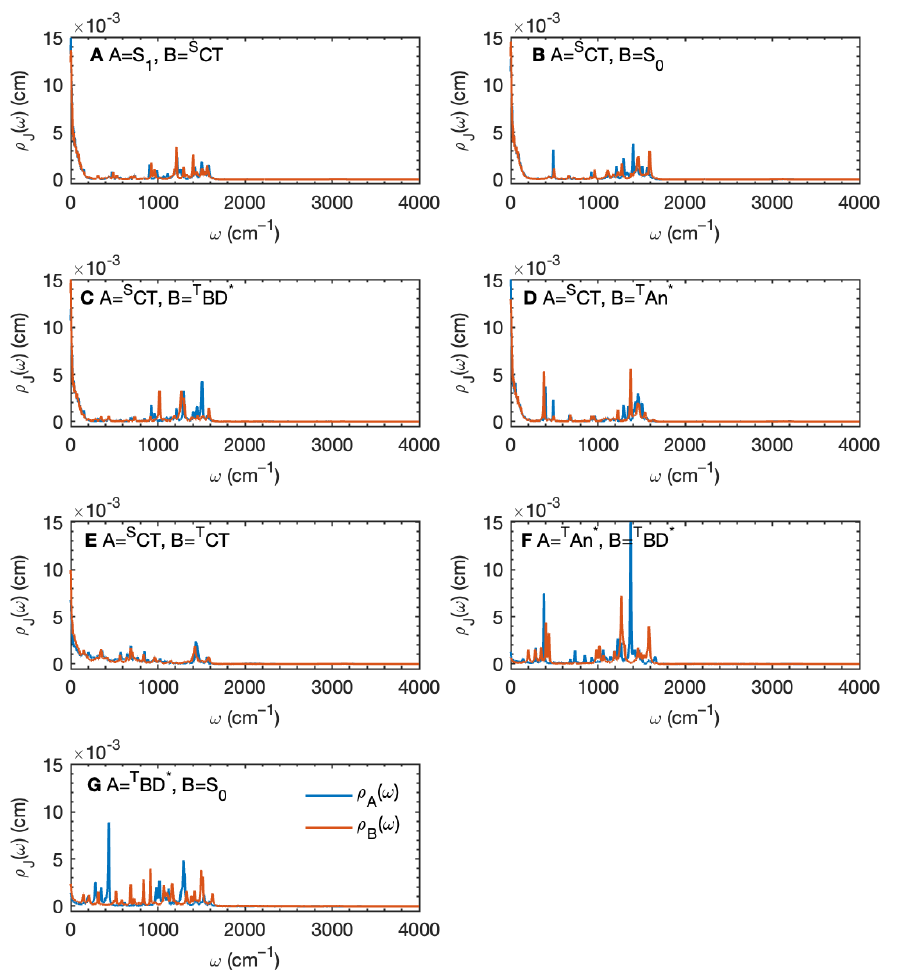}
    \caption{Spectral densities for the processes considered in this work.}
    \label{fig-spec-dens}
\end{figure*}

The spectral distributions for the various photophysical processes are given in Fig.~\ref{fig-spec-dens}.

\bibliography{si-bib.bib}
